\documentclass[]{aa} 
\usepackage{graphicx}
\usepackage{txfonts}
\begin{document}
%
\def\hi {H\,{\sc i}}
\def\hii {H\,{\sc ii}}
\def\water {H$_2$O}
\def\meth {CH$_{3}$OH}
\def\dg{$^{\circ}$}
\def\kms{km\,s$^{-1}$}
\def\ms{m\,s$^{-1}$}
\def\jyb{Jy\,beam$^{-1}$}
\def\mjyb{mJy\,beam$^{-1}$}
\def\solmass {\hbox{M$_{\odot}$}}
\def\solum {\hbox{L$_{\odot}$}} 
\def\d {$^{\circ}$}
\def\n {$n_{\rm{H_{2}}}$}
\def\kmsg{km\,s$^{-1}$\,G$^{-1}$}
\title{EVN observations of 6.7~GHz methanol maser polarization in massive star-forming regions.}

\author{G.\ Surcis \thanks{Member of the International Max Planck Research School (IMPRS) for Astronomy and Astrophysics at 
the Universities of Bonn and Cologne.}  \inst{, 1,2}
  \and 
  W.H.T. \ Vlemmings \inst{1,3}
 \and
  H.J. \ van Langevelde \inst{2,4}
  \and
  B. \ Hutawarakorn Kramer \inst{5,6}
  }

\institute{Argelander-Institut f\"{u}r Astronomie der Universit\"{a}t Bonn, Auf dem H\"{u}gel 71, 53121 Bonn, Germany\\
 \email{gsurcis@astro.uni-bonn.de}
 \and
 Joint Institute for VLBI in Europe, Postbus 2, 7990 AA Dwingeloo, The Netherlands
 \and
 Chalmers University of Technology, Onsala Space Observatory, SE-439 92 Onsala, Sweden
 \and
 Sterrewacht Leiden, Leiden University, Postbus 9513, 2300 RA Leiden, The Netherlands
 \and
 Max-Planck Institut f\"{u}r Radioastronomie, Auf dem H\"{u}gel 69, 53121 Bonn, Germany
 \and
 National Astronomical Research Institute of Thailand, Ministry of Science and Technology, Rama VI Rd., Bangkok 10400, Thailand
  }

\date{Received ; accepted}
\abstract
{The role of magnetic fields in the formation of high-mass stars is still under debate, and recent measurements of their orientation 
and strength by using polarized maser emissions are contributing new insights. Masers polarization, in particular of the 
6.7-GHz methanol masers, are one of the best probes of the magnetic field morphologies around massive protostars.  
}
{
Determining the magnetic field morphology around an increasing number of massive protostars at milliarcsecond resolution by observing 
6.7-GHz methanol masers is crucial to better understand the role of magnetic fields in massive star formation. 
} 
{The First EVN Group consists of 4 massive star-forming complexes: W51, W48, IRAS\,18556+0138, and W3(OH). These contain 
well-studied \hii ~regions from some of which molecular bipolar outflows were also detected (W51--e2, G35.20-0.74N).
Nine of the European VLBI Network antennas were used to measure the linear polarization and Zeeman-splitting of the 6.7-GHz 
methanol masers in the star-forming regions of the First EVN Group.}
{We detected a total of 154 \meth ~masers, one third of these towards W3(OH). 
Fractional linear polarization (1.2--11.5\%) was detected towards 55 \meth ~masers. The linear polarization vectors are well-ordered in all
the massive star-forming
regions. We measured significant Zeeman-splitting in 3 massive star-forming regions (W51, W48, and W3(OH)) revealing a range of
separations
$-3.5$~\ms$<\Delta V_{\rm{z}}<3.8$~\ms ~with the smallest $|\Delta V_{\rm{z}}|=0.4$~\ms. We were also able to compare our magnetic 
field results with those obtained from submillimeter wavelength dust observation in W51 and show that the magnetic field at low and high
resolutions are in perfect agreement.}
{}

\titlerunning{EVN group: magnetic field.}
\authorrunning{Surcis et al.}

\maketitle
\section{Introduction}
A number of different scenarios have been proposed to explain the formation of stars with masses larger than about 8~\solmass. 
These include formation through the merger of less massive stars (\textit{Coalescence model}; e.g., Bally \& Zinnecker \cite{bal05})  or 
through the accretion of unbound gas from the molecular cloud (\textit{Competitive accretion model}; e.g., Bonnell et al. \cite{bon04}). 
In the third scenario, the \textit{Core accretion model}, massive stars form through gravitational collapse, which involves disc-assisted accretion
 to overcome radiation pressure (e.g., McKee \& Tan \cite{mck03}). This scenario is similar to the favored picture of low-mass
star formation, in which magnetic fields are thought to regulate the collapse, to transfer the angular momentum and to power the bipolar outflows.
 Moreover, during the formation of low-mass stars the gravitational collapse of molecular clouds proceeds preferentially along the magnetic 
field lines, giving rise to large rotating disc or torus structures orthogonal to the magnetic field (e.g., Matsumoto \& Tomisaka \cite{mat04}).
 Consequently, the molecular bipolar outflows, which originate from the protostar, are driven parallel to the original magnetic field
orientation. Although the magnetic fields play such an important and crucial role in the formation of low-mass stars, their role in the 
formation of high-mass stars is still under debate. Fairly recently magnetic fields have been included in the theoretical simulations 
of high-mass star formation (e.g. Banerjee \& Pudritz \cite{ban07}). Therefore, providing new measurements of magnetic fields orientation and strength 
at milliarcsecond (mas) resolution is fundamental to understand the formation process of high-mass stars. \\
\indent  
Current 
dust observations of magnetic fields in massive star-forming regions are often limited to low density regions and/or envelopes at scales of several 
thousands astronomical units (e.g., Koch et al. \cite{koc10}). Linear polarization observations of dust also only provide information on the 
magnetic field in the plane of the sky, so these observations have been yet unable to probe the strength and the full structure of the magnetic 
field close to the protostars and around protostellar discs (e.g., Tang et al. \cite{tan10}).\\
\indent The best probes of magnetic fields in the high density regions close to massive protostars currently available are masers. Their bright 
and narrow spectral line emission is ideal for measuring the Zeeman-splitting even though the exact proportionality between the measured splitting and 
the magnetic field strength is still uncertain (see Vlemmings et al. \cite{vle11}), as well as for determining the orientation
 of the magnetic field in 3-dimension (e.g., Vlemmings et al. \cite{vle10}). Recently 6.7-GHz \meth ~masers, which are the most abundant maser 
species in massive star-forming regions, have started to play a crucial role in determining the magnetic field morphology close to 
massive protostars (e.g., Vlemmings et al. \cite{vle06b, vle10}, Surcis et al. \cite{sur09, sur11a}). The \meth ~masers probe magnetic fields along outflows 
and on surfaces of tori (Surcis et al. \cite{sur09};Vlemmings et al. \cite{vle10}; Surcis et al. \cite{sur11a}). Therefore, determining the magnetic field 
morphology around a large number massive protostars by observing the polarized emission of 6.7-GHz \meth ~masers is of fundamental importance. Here we show
 the results of our First EVN Group composed of 4 massive star-forming regions, which are described in details in Sect.~\ref{FES}. Observations 
of a second group of 5 sources are in the process of being analyzed. 

\section{The First EVN Group}
\label{FES}
We selected a first group of 4 massive star-forming regions among the northern hemisphere sources observed with the Effelsberg 100-m telescope where
6.7-GHz \meth ~maser Zeeman-splitting was measured (Vlemmings \cite{vle08}). The sources were selected based on their peak flux density
to allow potential detection of Zeeman-splitting in several individual \meth ~maser features.
\subsection{W51}
The W51 complex is one of the most luminous star-forming regions in the first quadrant of the Galactic plane. The high luminosity comes from a large
 number of O-type stars that are within the molecular cloud (Bieging \cite{bie75}; Carpenter \& Sanders \cite{car98}). Kundu \& Velusamy (\cite{kun67})
 found that W51 is divided in four different regions: W51A, W51B, W51C, and W51D. These regions are composed of several sub-regions, for 
instance W51A consists of G49.4-0.3 and G49.5-0.4 (e.g., Kang et al. \cite{kan10}). Moreover, G49.5-0.4 comprises the sources 
W51-Main/South (it contains the UC\hii ~regions W51--e1 and W51--e2; e.g., Gaume et al. \cite{gau93}), W51--IRS1, and W51--IRS2 (it 
contains W51--d2; e.g., Lacy et al. \cite{lac07}). \\

\noindent\textbf{W51--e1/e2}\\
\noindent W51--e1 and W51--e2 ($V_{\rm{lsr}}=59$~\kms) are the brightest molecular cores located at a distance of 5.41$^{+0.31}_{-0.28}$~kpc 
(Sato et al. \cite{sat10}) in the eastern edge of W51. Their masses are $\sim$150~\solmass ~and $\sim$110~\solmass, respectively (Zhang \& Ho \cite{zha97}). 
Observations of molecular lines showed evidence for infalling, or accreting, gas
 with a possible rotation around W51-e2 (e.g., Zhang et al. \cite{zha98}; Sollins et al. \cite{sol04}; Keto \& Klaassen \cite{ket08}). 
Keto \& Klaassen (\cite{ket08}) suggested a possible bipolar outflow along the North-West and South-East direction from 
W51-e2 (PA$\approx150$\d), which appears to be along the rotation axis of the ionized disc. \\
\indent OH maser polarization observations reveal that W51--e2 has two Zeeman pairs implying a magnetic field strength of
about $-$20~mG (Argon et al. \cite{arg02}). Vlemmings (\cite{vle08}) measured a Zeeman-splitting 
of the 6.7-GHz \meth~maser  $\Delta V_{\rm{Z}}=0.72\pm0.04$~\ms. Submillimeter Array (SMA) observations revealed that the inferred $B_{\perp}$ morphology is hourglass-like 
near the collapsing core of W51--e2, with its pinched direction parallel to the direction of the ionized accretion (Tang et al. \cite{tan09}).\\

\noindent\textbf{W51--IRS2}\\
\noindent W51--IRS2 is the most luminous massive star-forming region in the Milky Way (Erickson \& Tokunaga \cite{eri80}) and it is located at a distance of 
$5.1^{+2.9}_{-1.4}$~kpc (Xu et al. \cite{xu09a}). W51--IRS2 hosts a young O5-type protostar 
and an UC\hii ~region called W51--d2 (e.g., Gaume et al. \cite{gau93}; Zapata et al. \cite{zap08}). 
W51--d2 is located West w.r.t. the \water ~maser complex called W51--North at a projected distance of about 17000~AU. The source is associated 
with NH$_3$ and \meth ~masers
  as well as \water ~masers (Gaume et al. \cite{gau93}; Minier et al. \cite{min01}; Eisner et al. \cite{eis02}; Phillips \& van Langevelde \cite{phi05}). 
\subsection{W48}
The W48 complex is located in the Carina-Sagittarius spiral arm at a distance of 3.27$^{+0.56}_{-0.42}$~kpc (Zhang et al. \cite{zha09}). Onello et al. (\cite{one94})
showed that W48 consists of five \hii~regions:  W48A-E. W48A hosts a cometary UC\hii~region called G35.2--1.74 
 that is believed to be a site of massive star formation (Wood \& Churchwell \cite{woo89}; Roshi et al. \cite{ros05}), and two infrared sources IRS\,1 \
and IRS\,2 (Zeilik \& Lada \cite{zei78}). \meth ~masers are associated with G35.2--1.74 (Caswell et al. \cite{cas95}; Minier et al. \cite{min00}). The \meth ~maser 
region shows a ring-like structure of $\sim200\times400$~mas along a direction 
North-South.
The mid-infrared source 
MIR\,3 corresponds to G35.2--1.74 and it is quite predominant in the mid-infrared images of De Buizer et al. (\cite{deb05}). MIR\,1 coincides with the OH and 
\water ~masers site. \\
\indent Vlemmings (\cite{vle08}) measured a Zeeman-splitting of the \meth ~masers of 0.32$\pm$0.02~\ms. 
Submillimeter polarimetry observations with the
Submillimetre Common User Bolometer Array (SCUBA) revealed two cores, the main and bright core in the East, W48main, and the small core in the 
West, W48W (Curran et al. \cite{cur04}). The two cores appear to be connected by a ridge of dust and gas. The degree of polarization is lower (0.5\%) towards the 
intensity peak of W48main, which is close to G35.2--1.74, and is higher towards the edges. Moreover, the magnetic field orientation is complicated and near the \meth ~maser
region, towards the centre of W48main, the magnetic field has a NW-SE orientation. Whereas, in W48W the magnetic field has an ordered orientation 
North-South. This changing in the orientation has been explained by twisting of the magnetic field towards the centre of W48main (Curran et al. \cite{cur04}).
\subsection{IRAS\,18556+0138}
IRAS\,18556+0138 is at a distance of 2.19$^{+0.24}_{-0.20}$~kpc (Zhang et al. \cite{zha09}) and has a systemic velocity $V_{\rm{sys}}\sim34$~\kms 
~(Matthews et al. \cite{mat84}). This IRAS source is associated with the massive star-forming region G35.2--0.74N that contains an embedded B0.5 
star  with a luminosity of $\sim10^4$~\solum ~and a stellar mass of $\sim15$~\solmass ~(Dent et al. \cite{den85}; Gibb et al. \cite{gib03}). 
A collimated CO--molecular bipolar 
outflow (PA$=58$\d, Gibb et al. \cite{gib03}) was detected from G35.2--0.74N (e.g., Dent et al. \cite{den85}). The outflow has a size of about 2~pc
and its blue- ($25.3$~\kms$<V_{\rm{outf-IRAS}}^{\rm{blue}}<30.3$~\kms) and red-shifted ($37.8$~\kms$<V_{\rm{outf-IRAS}}^
{\rm{red}}<43.8$~\kms) parts are located South-West and North-East w.r.t. the protostar (L\'{o}pez-Sepulcre et al. \cite{lop09}), respectively. A
radio/infrared jet was detected with a position angle PA$\sim0$\d (Fuller et al. \cite{ful01};
De Buizer \cite{deb06}) suggesting that the outflow is driven by the precessing jet (e..g.; Mart\'{i} et al. \cite{mar93}). 
In addition, a flattened structure rotating perpendicularly (PA$=148$\d) to the outflow axis was also observed (e.g.,
 L\'{o}pez-Sepulcre et al. \cite{lop09}). L\'{o}pez-Sepulcre et al. (\cite{lop09}) measured a velocity 
range of the rotating structure between 33.45~\kms ~(NW) and 34.75~\kms ~(SE), and a SE-NW size of $\sim30''$, which at 2.19~kpc correspond to about 66000~AU 
($\sim0.3$~pc). \\
\indent Hutawarakorn \& Cohen (\cite{hut99}) measured magnetic field strengths between --2.5~mG, North of G35.2--0.74N, and +5.2~mG, South by studying the 
Zeeman-splitting of the OH masers. They also found an orientation of the magnetic field parallel to the outflow, even if Faraday rotation could play an important role. 
Using Effelsberg observations Vlemmings (\cite{vle08}) measured a Zeeman-splitting of the \meth ~maser emission $\Delta V_{\rm{Z}}=0.81\pm0.04$~\ms.
\subsection{W3(OH)}
W3(OH) is the most studied UC\hii ~region in our Galaxy. It is located in the Perseus spiral arm at a distance of 1.95$\pm$0.04~kpc (Xu et al. \cite{xu06}).
Infrared and radio continuum observations indicated that the central ionizing star is a O7 star with a mass of about 30~\solmass ~(Scott \cite{sco81};
Campbell et al. \cite{cam89}).
Moscadelli et al. (\cite{mos99,mos10}) derived accurate proper motion of the 12-GHz \meth ~masers of which the most intense concentrate in a small area
 towards the North (called the \textit{northern clump}) of the UC\hii ~region. A second clump is located South, and between the two clumps an 
isolated single spot maser has been found. This North-South filamentary structure of 12-GHz \meth ~masers is also shown by the 6.7-GHz \meth ~and 
the OH masers (Harvey-Smith \& Cohen \cite{har06}). The brightest 12-GHz \meth ~masers of the northern clump are distributed along an axis at PA$=141$\d ~and show a regular 
trend of radial velocities with position, suggesting that they could trace a self-gravitating, low-mass ($M\approx1.5$~\solmass)
circumstellar disk ($R\approx270$~AU)
in the phase of being photo-evaporated by the strong UV-radiation field escaping from the W3(OH) UC\hii ~region (Moscadelli et al. \cite{mos10}). 
6.7-GHz \meth ~masers were detected for the first time at high angular resolution by Menten et al. (\cite{men92}) and they coincide with the masers at 
12-GHz (Moscadelli et al. \cite{mos99}). In particular some 6.7-GHz \meth ~masers trace a similar linear structure as shown by the 12-GHz \meth ~masers (PA$=130$\d; 
Harvey-Smith \& Cohen \cite{har06}). \\
\indent Polarization observations of the \meth ~masers with MERLIN showed that the magnetic field is parallel to the extended filament but with a 
complex structure in the dominating northern clump (Vlemmings et al. \cite{vle06b}). Wright et al. (\cite{wri04}) detected linear polarization from
the 1.6-GHz OH maser emission, showing a wider spread maser polarization angles, with the median 
polarization angle $\langle \chi_{\rm{med}}^{\rm{OH}} \rangle=104\pm27$\d, that has been attributed to Faraday rotation along the maser path.
A Zeeman-splitting of the 6.7-GHz \meth ~masers $\Delta V_{\rm{Z}}=0.141\pm0.003$~\ms ~was measured by Vlemmings (\cite{vle08}). 

\section{Observations and data reduction}\label{obssect}
The group was observed at 6.7-GHz in full polarization spectral mode with nine of the European VLBI Network\footnote{The European VLBI Network
is a joint facility of European, Chinese, South African and other radio astronomy institutes funded by their national research councils.}
 antennas (Jodrell2, Cambridge, Effelsberg, Onsala, Medicina, Torun, Noto, Westerbork, and Yebes-40\,m), for a total 
observation time of 20~h, on November 2 (W48, IRAS\,18556+0138, and W3(OH)) and 3 (W51), 2009 (program codes ES063A and ES063B). The bandwidth was 
2~MHz, providing a velocity range of $\sim100$~\kms. The data were hardware-correlated with the Joint Institute for VLBI in Europe (JIVE) correlator
(Schilizzi et al. \cite{sch01}) using 1024 channels and generating all 4 polarization 
combinations (RR, LL, RL, LR) with a spectral resolution of 1.9~kHz ($\sim$0.1~\kms). In order to detected \meth ~maser towards both W51--e1/e2
and W51--IRS2 the data of W51 were correlated twice with two different correlator positions. The correlator position of W51--e1/e2 was the 
pointing position of the observations.\\
\indent The data were edited and calibrated using AIPS. The bandpass, the delay, the phase, and the polarization calibration were 
performed on the calibrators J2202+4216 (W48 and W51) and 3C286 (IRAS\,18556+0138 and W3(OH)). Fringe-fitting and self-calibration were performed 
on the brightest maser feature of each star-forming region. Then the \textit{I}, \textit{Q}, \textit{U}, \textit{RR}, and \textit{LL} 
cubes were imaged (2~arcsec~$\times$~2~arcsec, rms ranges from $4$~\mjyb ~to $50$~\mjyb depending on the source) using the AIPS task IMAGR. The 
beam-sizes were 5.5~mas~$\times$~4.5~mas for W51,
6.8~mas~$\times$~5.4~mas for W48, 7.5~mas~$\times$~4.8~mas for IRAS\,18556+0318, and 8.7~mas~$\times$~4.1~mas for W3(OH).
The \textit{Q} and \textit{U} cubes were combined to produce cubes of polarized intensity ($P_{\rm{l}}=\sqrt{Q^{2}+U^{2}}$) and polarization angle 
($\chi=1/2\times atan(U/Q)$). The EVN observations were 
obtained close to VLA polarization observations\footnote{http://www.aoc.nrao.edu/$\sim$~smyers/calibration/} made by the NRAO at the beginning of November
 2009, during which the polarization angles of J2202+4216 and 3C286 were $\sim-29$\d$\!\!.2$ and $\sim11$\d$\!\!.5$ respectively. Hence, we were able to
estimate the polarization angles with a systemic error of no more than $\sim$~3\d. The formal error on $\chi$ are due to thermal noise. This error is given
by $\sigma_{\chi}=0.5 ~\sigma_{P}/P \times 180^{\circ}/\pi$ (Wardle \& Kronberg \cite{war74}), where $P$ and $\sigma_{P}$ are the polarization intensity
and corresponding rms error respectively.
 We were able to obtain the absolute position of the brightest features of W51--e2, W51--IRS2, and W3(OH) through fringe rate mapping using the AIPS 
task FRMAP. The absolute position errors were less than 30~mas.
\section{Analysis}
\indent We identified the \meth ~maser features using the process described in Surcis et al. (\cite{sur11a}, \cite{sur11b}). The maser 
features that showed linear
polarization emission were fitted by using the adapted full radiative transfer method code for \meth ~masers described in Surcis et al. (\cite{sur11a}) 
and based on the models for \water ~masers of Nedoluha \& Watson (\cite{ned92}).   The fit provides the emerging brightness temperature 
($T_{\rm{b}}\Delta\Omega$) and the intrinsic thermal linewidth ($\Delta V_{\rm{i}}$). We modeled the observed linear polarized and total intensity maser 
spectra by gridding $\Delta V_{\rm{i}}$ from 0.5 to 1.95~\kms. Considering $T_{\rm{b}}\Delta\Omega$ and $P_{\rm{l}}$ we 
determined the angles between the maser propagation direction and the magnetic field ($\theta$), if $\theta>\theta_{\rm{crit}}=55$\d, where $\theta_{\rm{crit}}$
is the Van Vleck angle, the magnetic field appears 
to be perpendicular to the linear polarization vectors, otherwise it is parallel.
 We determined the Zeeman-splitting ($\Delta V_{\rm{Z}}$) from the cross-correlation between the RR and LL spectra, like was successfully used in 
Surcis et al. (\cite{sur09}, 
\cite{sur11b}) for the polarized \meth ~maser emission detected in W75N and NGC\,7538. The dynamic range of the RR and LL cubes decreases close 
to the strongest maser emission of each group because of the residual calibration errors. As a result, we were not able to determine 
$\Delta V_{\rm{Z}}$ for any \meth ~maser features of W3(OH) with a peak flux less than $\sim$15~\jyb.
\begin{figure*}[th!]
\centering
\includegraphics[width = 7 cm]{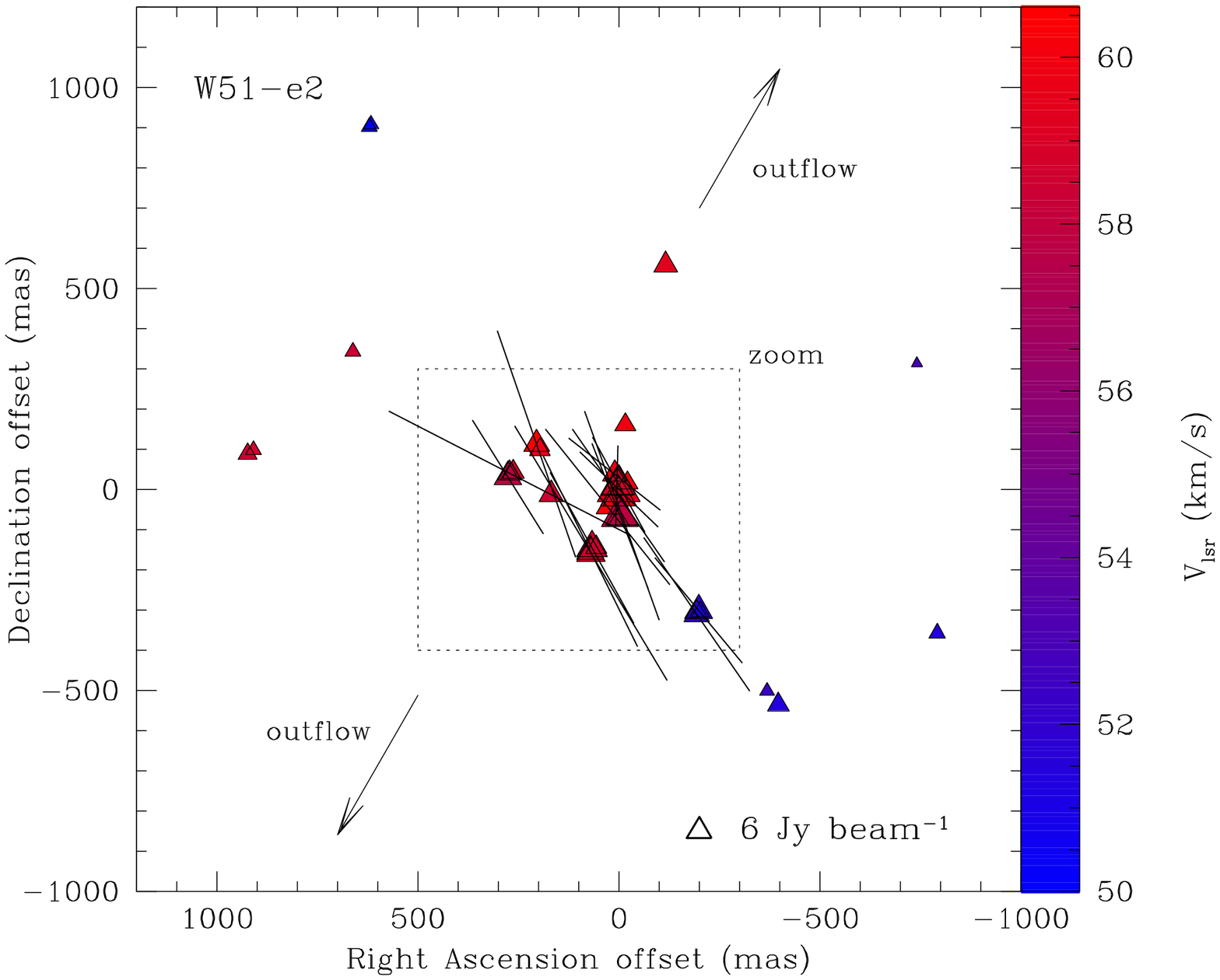}
\includegraphics[width = 7 cm]{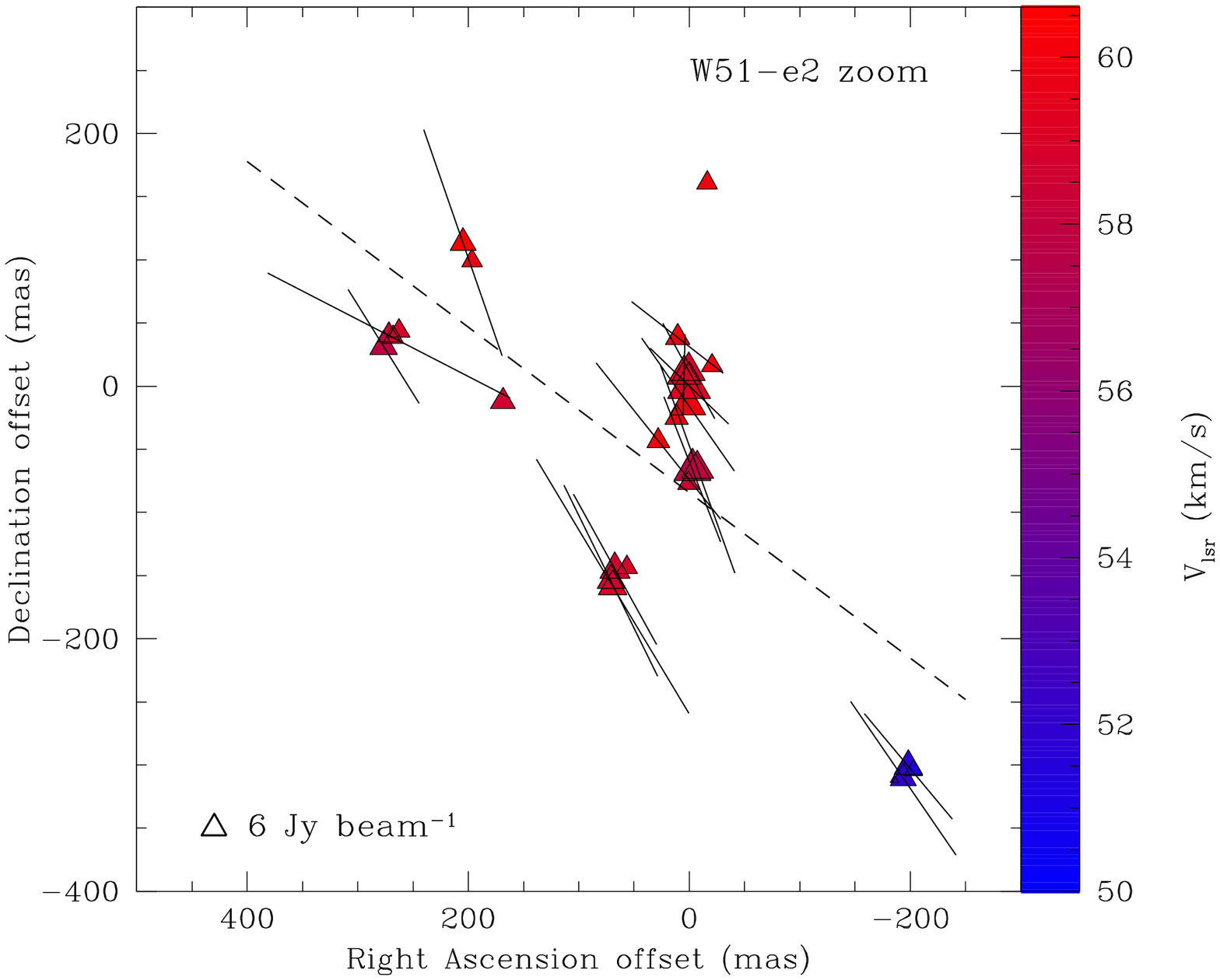}
\caption{Left panel: a view of the \meth ~maser features detected around W51--e2. Right panel: a zoom-in view of the boxed region of the
left panel. The triangles symbols are the identified \meth ~maser features scaled logarithmically according to their peak
flux density (Table~\ref{WE_tab}). The maser LSR radial velocity is indicated by color. A 6~\jyb ~symbol is plotted for illustration in both
panels. The linear polarization vectors, scaled logarithmically according to polarization fraction $P_{\rm{l}}$, are overplotted. The two arrows
indicate the direction of the bipolar outflow (PA$\sim150$\d; Keto \& Klaassen \cite{ket08}). The dashed line is the best linear fit of the \meth ~maser features
 in the right panel (PA$=57$\d)}
\label{W51_e2cp}
\end{figure*}
\section{Results}
\label{res7}
In Tables~\ref{WE_tab}--\ref{W3OH_tab} (see online material) we list all the 154 \meth ~maser features detected towards the 4 massive star-forming regions
observed with the EVN. The description of the maser distributions and the polarization results are reported for each source separately in
Sects.~\ref{W51_res}--\ref{W3OH_res}.
\subsection{\object{W51}}
\label{W51_res}
We detected 6.7-GHz \meth ~maser emission in the star-forming region W51--e1/e2 and W51--IRS2. In particular, 37 features towards W51--e2 
(named as W51E.01-W51E.37 in Table~\ref{WE_tab}) and 18 towards W51--IRS2 (named as W51N.01-W51N.18 in Table~\ref{WN_tab}). No \meth ~maser emission
(8$\sigma=0.1$~\jyb) was detected around W51--e1.\\

\noindent\textbf{W51--e2}\\
\noindent In Fig.~\ref{W51_e2cp} we show the \meth ~maser features colour-coded according to the LSR radial velocities ($50$~\kms$<V_{\rm{lsr}}<61$~\kms).
The maser features near the 870~$\rm{\mu}$m continuum peak (Tang et al. \cite{tan09}) lie in an elongated structure (PA$=57$\d)
perpendicular to the molecular bipolar outflow (PA$\sim150$\d; Keto \& Klaassen \cite{ket08}) and close to the ionized accreting flow direction (PA$\sim30$\d; 
Keto \& Klaassen \cite{ket08}). This suggests that the masers potentially trace a disc structure perpendicular to the bipolar outflow. 
However, the velocities of the \meth ~maser features are consistent with the velocities at the centre of the H53$\alpha$ emission, which 
traces the rotating flow (Keto \& Klaassen \cite{ket08}). This suggests that the masers might probe material that is being accreted onto the disc itself, 
like in Cepheus~A and NGC7538 (Vlemmings et al. \cite{vle10}; Torstensson et al. \cite{tor11}; Surcis et al. \cite{sur11a}), rather than the disc material.\\
\indent We detected linearly polarized emission ($P_{\rm{l}}=1.2\%-4.0\%$) in 16 \meth ~maser features that are all located close to the centre 
of W51--e2. The error weighted linear polarization angle is $\langle\chi\rangle_{\rm{W51-e2}}=33$\d$\pm16$\d. The full radiative transfer method 
code for \meth ~masers was able to fit 14 features. The intrinsic thermal linewidths and the emerging brightness temperatures are given in 
column 9 and 10 of Table~\ref{WE_tab} and their weighted values are $\langle\Delta V_{\rm{i}}\rangle_{\rm{W51-e2}}=0.7^{+0.6}_{-0.2}$~\kms ~and 
$\langle T_{\rm{b}}\Delta\Omega\rangle_{\rm{W51-e2}}\approx10^{9}$~K~sr, respectively. The angles between the maser propagation direction and the magnetic
 field ($\theta$) are almost constant for all the \meth ~maser 
features. The feature W51E.17 shows $\theta=90$\d ~indicating that the feature could be partially saturated. Hence, excluding this feature, 
$\langle \theta\rangle_{\rm{W51-e2}}=79$\d$^{+11^{\circ}}_{-40^{\circ}}$. Since $\theta>\theta_{\rm{crit}}=55$\d ~the magnetic field is most 
likely perpendicular to the linear polarization vectors even though it might also be parallel because of the relatively large errors in column
12 of Table~\ref{WE_tab}. We measured Zeeman-splitting (column 11) in 4 \meth ~maser features. The weighted value is 
$\langle \Delta V_{\rm{Z}} \rangle_{\rm{W51-e2}}=-1.6\pm1.2$~\ms ~and in absolute value is twice larger than that measured by Vlemmings (\cite{vle08}).\\

\noindent\textbf{W51--IRS2}\\
We detected 18 \meth ~maser features with peak flux density between 0.16~\mjyb ~and 12~\mjyb ~and with velocities  
$54$~\kms$<V_{\rm{lsr}}<61$~\kms.
 The \meth ~maser features, all of which are associated with the UC\hii ~region d2, seem to lie in an arch structure 
oriented South-West. We detected linearly polarized emission only towards the brightest maser feature (W51N.11, $P_{\rm{l}}=1.3\%$ and 
$\chi_{\rm{W51-IRS2}}=27$\d$\pm13$\d). The intrinsic thermal linewidth and the emerging brightness temperature are in column 9 and 10 of 
Table~\ref{WN_tab} and the angle $\theta$ (column 13) indicates that the magnetic field is likely parallel to the linear polarization vector.
 Indeed, taking into account the errors, $\theta$ angles are less than $\theta_{\rm{crit}}=55$\d. No Zeeman-splitting was measured.
\begin{figure}[h!]
\centering
\includegraphics[width = 7 cm]{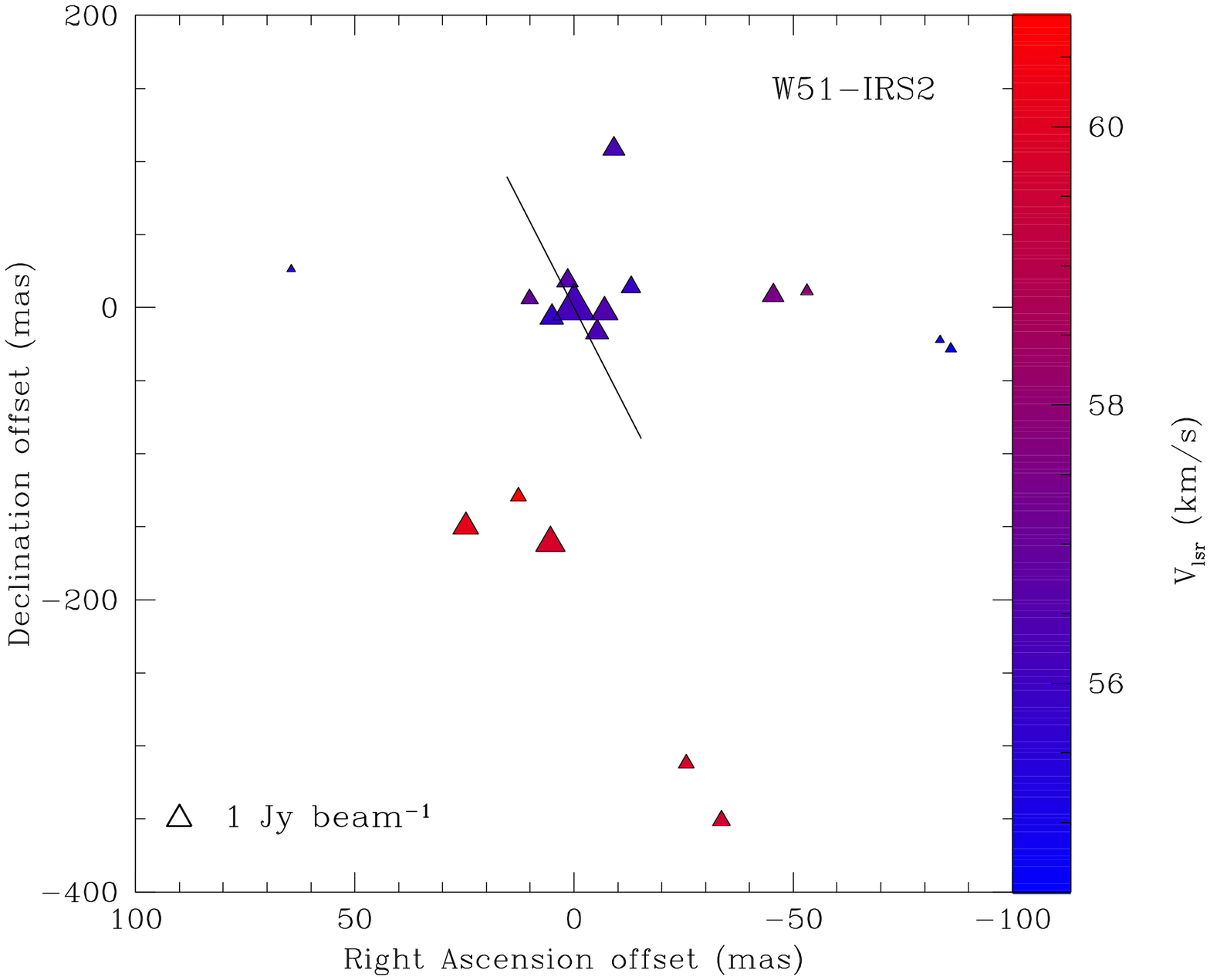}
\caption{A close-up view of the \meth ~maser features detected around W51--IRS2. The triangles symbols are the identified \meth ~maser features 
scaled logarithmically according to their peak
flux density (Table~\ref{WN_tab}). The maser LSR radial velocity is indicated by color. A 1~\jyb ~symbol is plotted for illustration.
 The linear polarization vectors, scaled logarithmically according to polarization fraction $P_{\rm{l}}$, are overplotted.}
\label{W51_IRS2cp}
\end{figure}
\begin{figure}[h!]
\centering
\includegraphics[width = 7 cm]{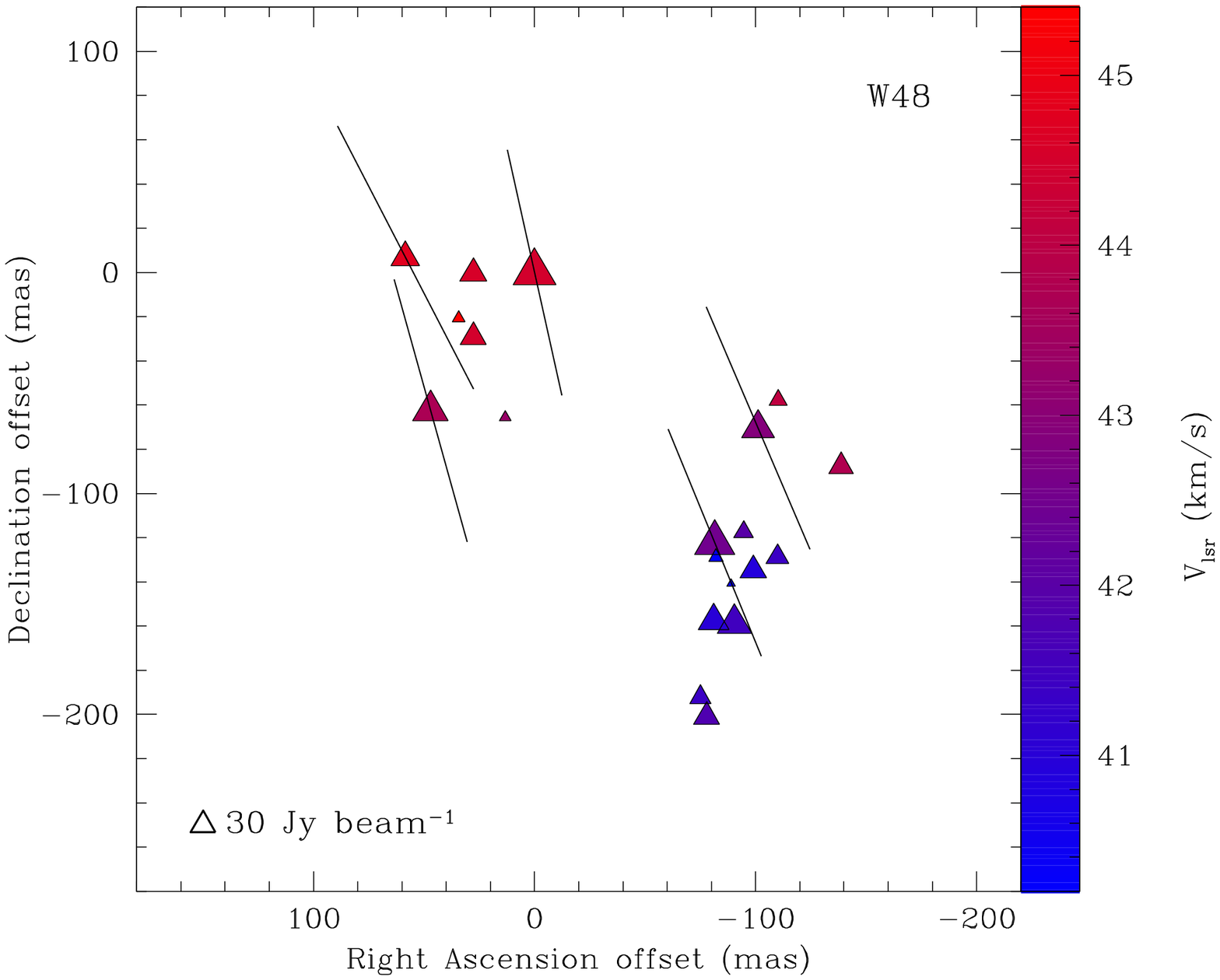}
\caption{A close-up view of the \meth ~maser features detected around W48 (G35.20--1.74). The triangles symbols are the identified maser 
features scaled logarithmically according to their peak
flux density (Table~\ref{W48_tab}). The maser LSR radial velocity is indicated by color. A 30~\jyb ~symbol is plotted for illustration.
  The linear polarization vectors, scaled logarithmically according to polarization fraction $P_{\rm{l}}$, are overplotted.}
\label{W48_cp}
\end{figure}
\subsection{\object{W48}}
\label{W48_res}
In Fig.~\ref{W48_cp} all the 20 \meth ~maser features are shown, named in Table~\ref{W48_tab} as W48.01--W48.20. The \meth ~maser features 
appear to be symmetrically distributed, with the red-shifted features ($V_{\rm{lsr}}>43$~\kms) located in the North-East and those blue-shifted
 ($V_{\rm{lsr}}<43$~\kms) in the South-West. All the \meth ~maser features are associated with the submillimetre--core W48main detected with 
SCUBA by Curran et al. (\cite{cur04}) and are closer to MIR3 than to MIR1 implying that the \meth ~maser features are not related to the other two maser species
(\water ~and OH).\\
\indent Linearly polarized emission was detected towards 5 \meth ~maser features that show very high linear polarization fraction 
($P_{\rm{l}}=4.4\%-6.5\%$). These high values of $P_{\rm{l}}$ suggest that the features might be partially saturated. Indeed the emerging
brightness temperatures obtained by using the full radiative transfer method code are (for all but W48.10) greater than the limit below which \meth 
~maser features can be considered unsaturated, i.e. $T_{\rm{b}}\Delta \Omega=2.6\times10^9$~K~sr (Surcis et al. \cite{sur11a}). As reported in
 Surcis et al. (\cite{sur11a}), 
the intrinsic thermal linewidth for saturated masers is overestimated so the observed linewidths $\Delta v_{\rm{L}}$ might be equal to 
$\Delta V_{\rm{i}}$ (column 6 and 9 of Table~\ref{W48_tab}, respectively). The $\theta$ angles (column 12) are all greater than 55\d ~so
the magnetic field is certainly perpendicular to the linear polarization vectors. Zeeman-splitting was measured only for the brightest feature 
(W48.14) and its value is $\Delta V_{\rm{Z}}=3.8\pm0.4$~\ms, which is 10 times larger than that measured by Vlemmings (\cite{vle08}).
\subsection{\object{IRAS\,18556+0318}}
\label{IRAS_res}
In Table~\ref{IRAS_tab}, named as I18556.01--I18556.28, we report all the identified 6.7-GHz \meth ~maser features. In particular, they are associated with the source G35.20-0.74N and they can be divided in two groups (A and 
\begin{figure}[h!]
\centering
\includegraphics[width = 7 cm]{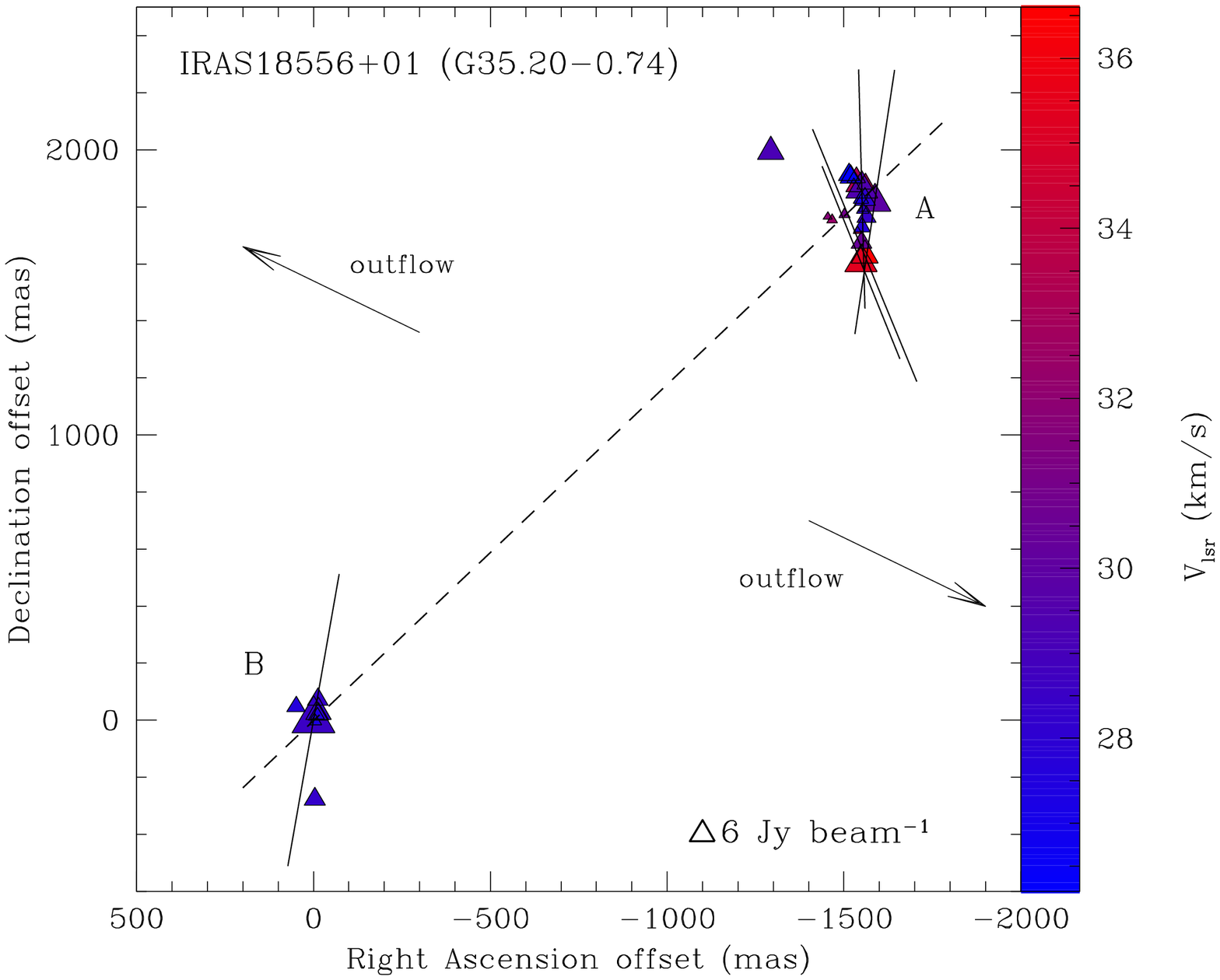}
\caption{A close-up view of the \meth ~maser features detected around IRAS\,18556+0138 (G35.20--0.74N). The triangles symbols are the identified 
\meth ~maser features scaled logarithmically according to their peak flux density (Table~\ref{IRAS_tab}). The maser LSR radial
 velocity is indicated by color. A 6~\jyb ~symbol is plotted for illustration. The linear polarization vectors, scaled 
logarithmically according to polarization fraction $P_{\rm{l}}$, are overplotted. The dashed line is the best linear fit of the \meth ~masers 
(PA$=140$\d). The two arrows indicate the direction, but not the origin, of the collimated CO-molecular bipolar outflow (PA$=58$\d; 
Gibb et al. \cite{gib03}).}
\label{IRAScp}
\end{figure}
B). Group A is composed of 20 \meth ~maser features (27.60~\kms~$<V_{\rm{lsr}}<$~36.56~\kms) that seem to be along an arch structure (top panel of 
Fig.~\ref{IRAScp_AB}), while Group B ($27$~\kms$<V_{\rm{lsr}}<29$~\kms) is located about 2.3~arcsec ($\sim$5000~AU at 2.19~kpc) South-East  
of group A and close to the peak of the radio continuum emission shown in De Buizer (\cite{deb06}). The maser distribution
coincides with that reported by Sugiyama et al. (\cite{sug08}). The two groups of \meth ~masers seem to be at the ends of a linear structure 
with PA$\sim140$\d and a radius of about 5000~AU, which is almost perpendicular to the direction of the CO-outflow (PA$=58$\d; Gibb et al. \cite{gib03}). 
This is suggested by the position angle of this linear structure (PA$=140$\d) that agrees with the position angles
obtained both from the OH maser emission (Hutawarakorn \& Cohen \cite{hut99}) and from the $\rm{CO^{18}O(2-1)}$ emission (L\'{o}pez-Sepulcre et al. \cite{lop09}).
If we consider the velocities of the \meth 
~and OH masers, of the large flattened structure, of the outflows, and of the source itself we find that the whole picture is quite different. 
The velocities of the two maser species are in good agreement, both are in the range $21$~\kms$<V_{\rm{masers}}<37$~\kms, and probably they are 
tracing the same gas. While the mean radial 
velocity of the large flattened structure is identical to the LSR velocity of the source ($V_{\rm{lsr}}^{G35.20}=34$~\kms, Hutawarakorn \& Cohen
\cite{hut99}). 
\begin{figure}[h!]
\centering
\includegraphics[width = 7 cm]{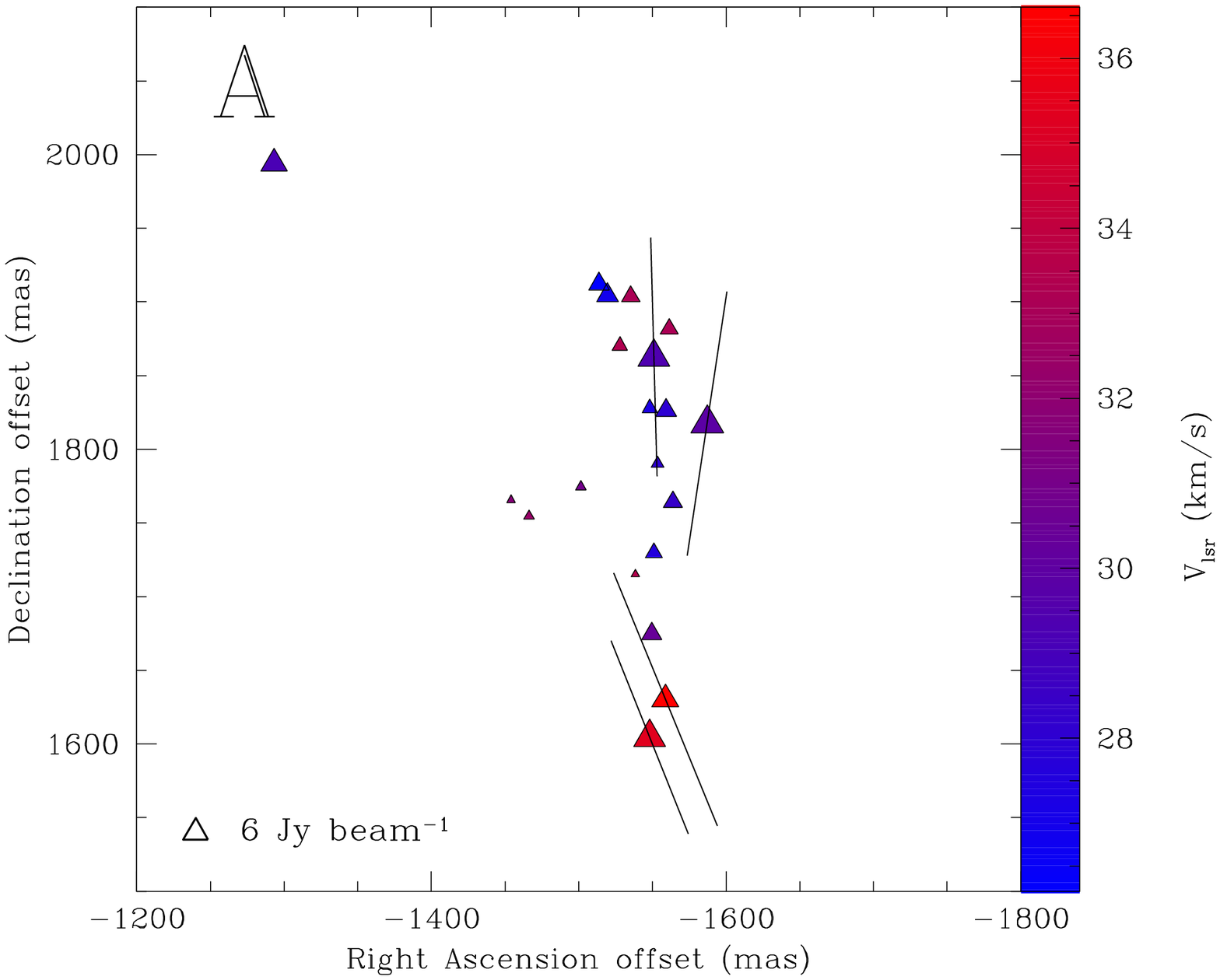}
\includegraphics[width = 7 cm]{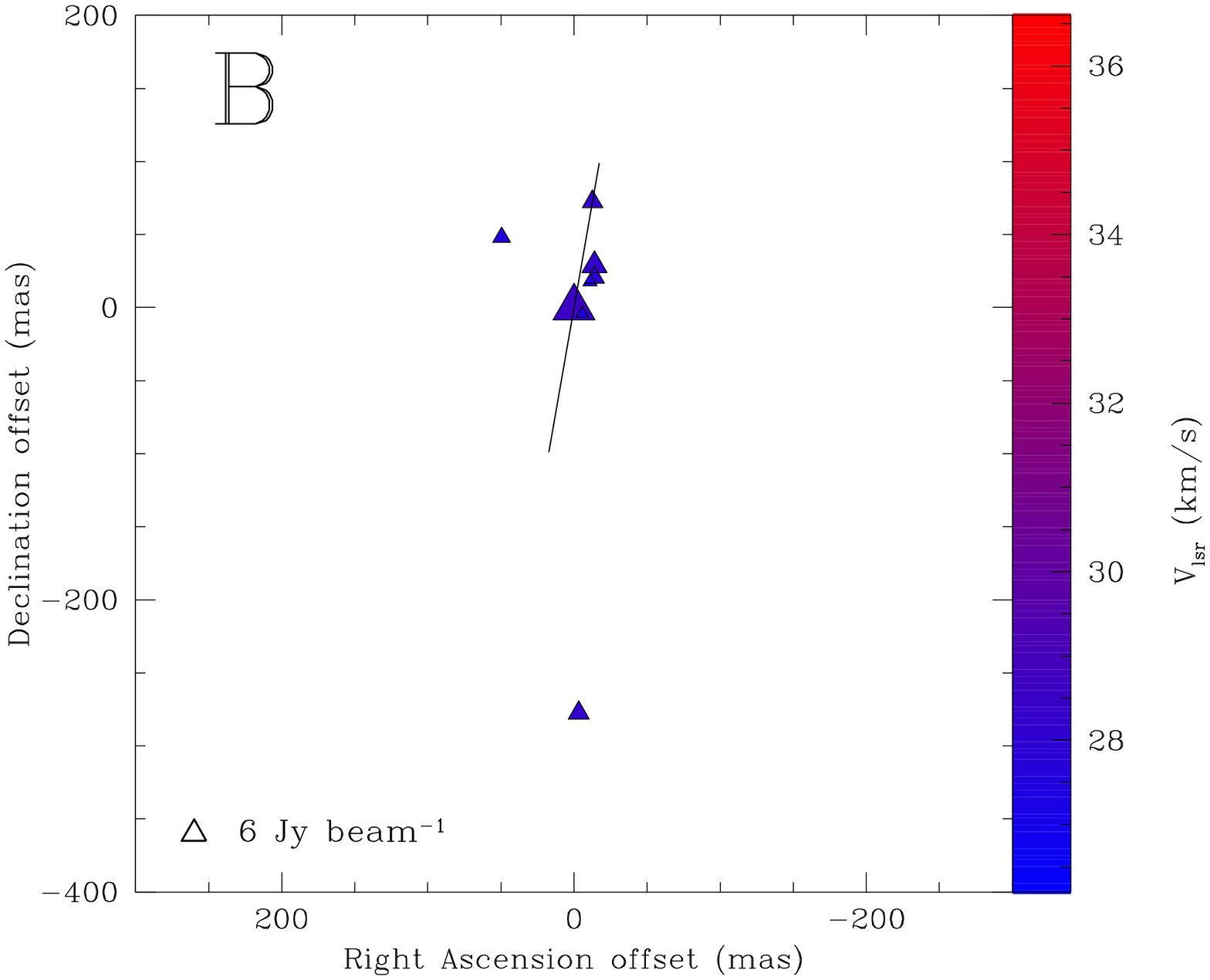}
\caption{Zoom-in view of the two \meth ~maser clusters of IRAS\,18556+0138 (G35.20--0.74N).}
\label{IRAScp_AB}
\end{figure}
The velocities of group B, which should be at the centre of the linear structure, are about 6~\kms ~blueshifted with respect to $V_{\rm{lsr}}^{G35.20}$ and 
 in the velocity range of the blue-shifted part of the outflow  ($25.3$~\kms$<V_{\rm{outf-IRAS}}^{\rm{blue}}<30.3$~\kms). This suggests that group B is 
associated with the outflow rather than with the linear structure. Also the velocities of group A appear to be in agreement with the outflow velocities
($25.3$~\kms$<V_{\rm{outf-IRAS}}<43.8$~\kms) although the arch-like distribution indicates an association 
with a torus structure. This torus structure might be related to the large flattened structure (33.45~\kms$<V_{\rm{torus-IRAS}}<$34.75~\kms).\\
\indent We detected linear polarization in 5 \meth ~maser features ($P_{\rm{l}}=4.8\%-11.5\%$) and the full radiative transfer method code was 
able to properly fit only two of them (I18556.08 and I18556.11). This is due to the likely high saturation degree of the \meth ~maser features. For the features I18556.01, I18556.04, 
and I18556.27 the gridding of $\Delta V_{\rm{i}}$ used for fitting the linear polarized and total intensity spectra was insufficient and consequently 
the code provided only an upper limit for $\Delta V_{\rm{i}}$. All the emerging brightness temperatures obtained from the fit, except I18556.11, reveal 
that the features have a high saturation degree. In particular, the reference maser feature I18556.27 shows the highest linear polarization fraction among 
all the 6.7-GHz \meth ~maser features detected in the whole EVN group ($P_{\rm{l}}=11.5\%$). In 
G35.20-0.74 the magnetic field is also perpendicular to the linear polarization vectors ($\theta>55$\d). No 6.7-GHz \meth ~maser Zeeman-splitting was measured in this 
source.

\subsection{\object{W3(OH)}}
\label{W3OH_res}
\begin{figure}[h!]
\centering
\includegraphics[width = 8 cm]{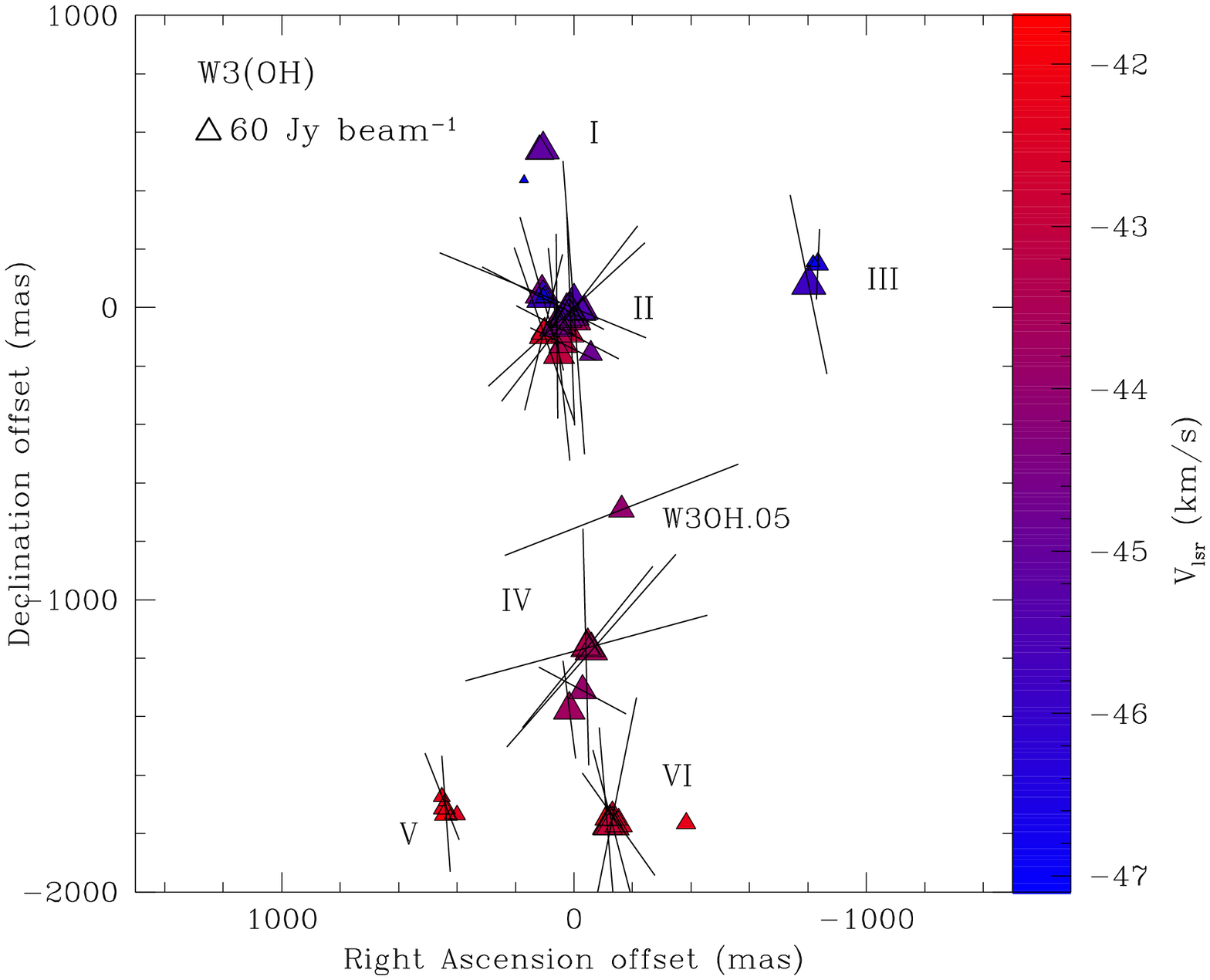}
\caption{A complete view of the \meth ~maser features detected around W3(OH). The triangles symbols are the identified 
\meth ~maser features scaled logarithmically according to their peak flux density (Table~\ref{W3OH_tab}). The maser LSR radial
 velocity is indicated by color. A 60~\jyb ~symbol is plotted for illustration. The linear polarization vectors, scaled 
logarithmically according to polarization fraction $P_{\rm{l}}$, are overplotted.}
\label{W3OHcp}
\end{figure}

\begin{figure*}[th!]
\centering
\includegraphics[width = 7 cm]{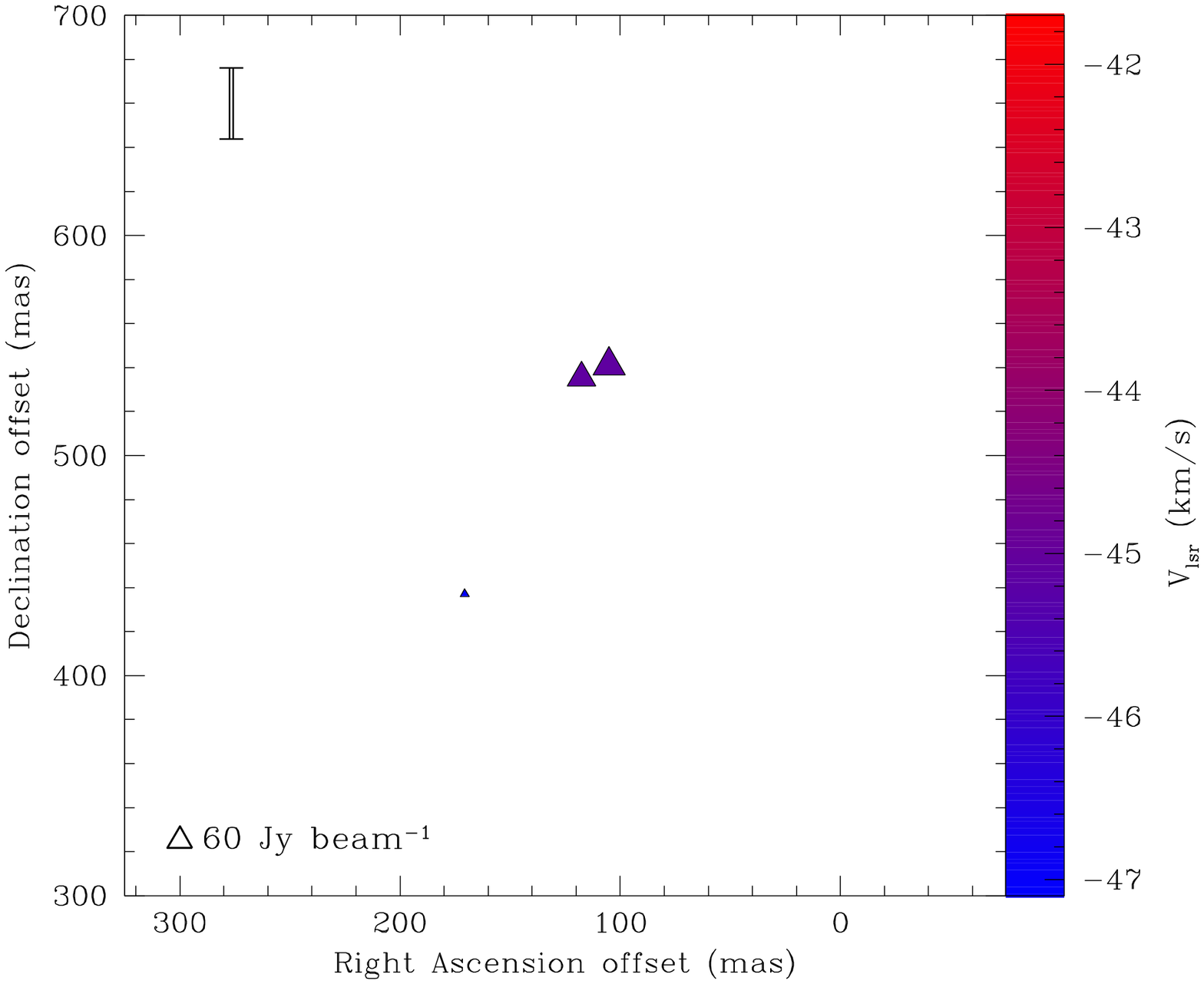}
\includegraphics[width = 7 cm]{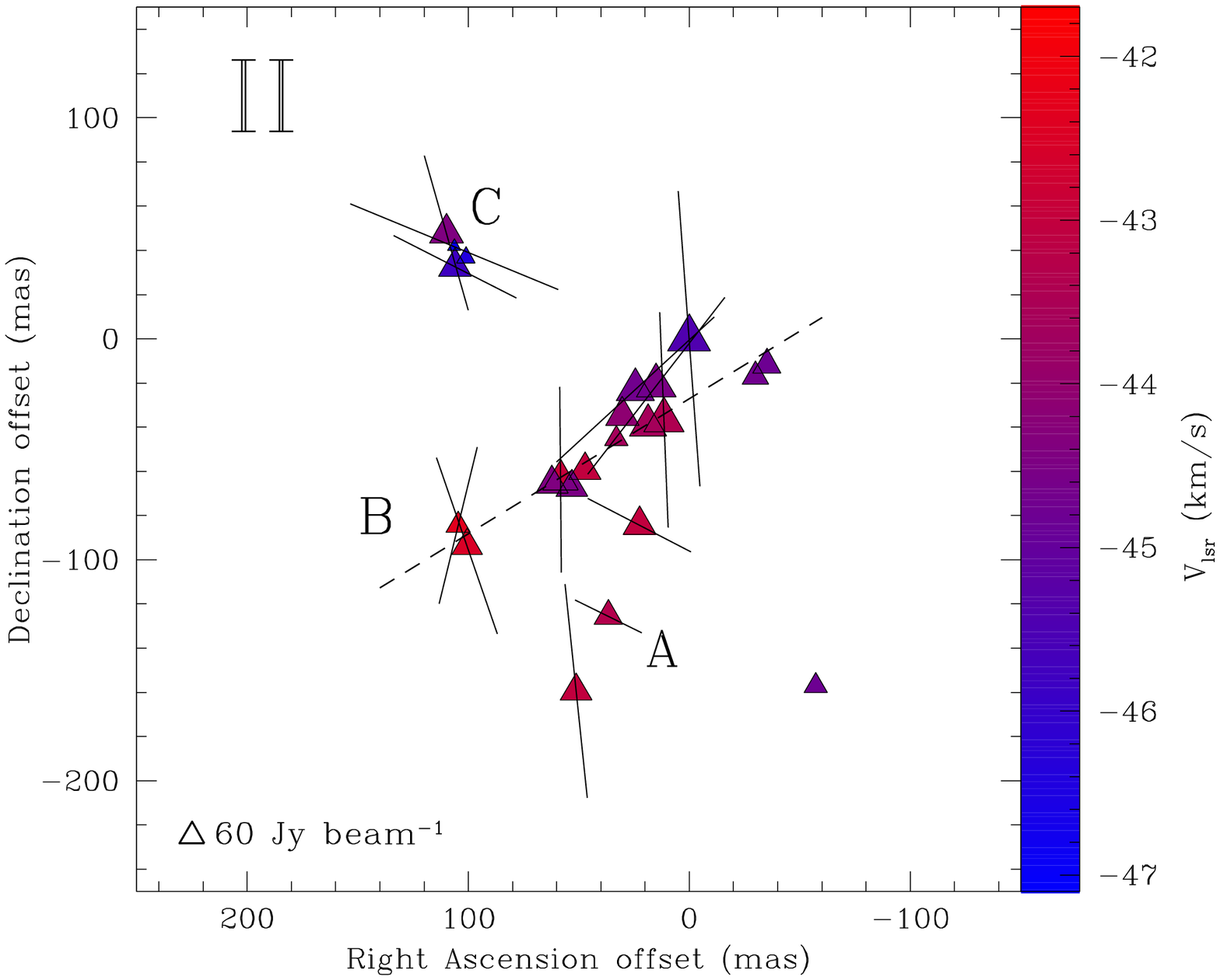}
\includegraphics[width = 7 cm]{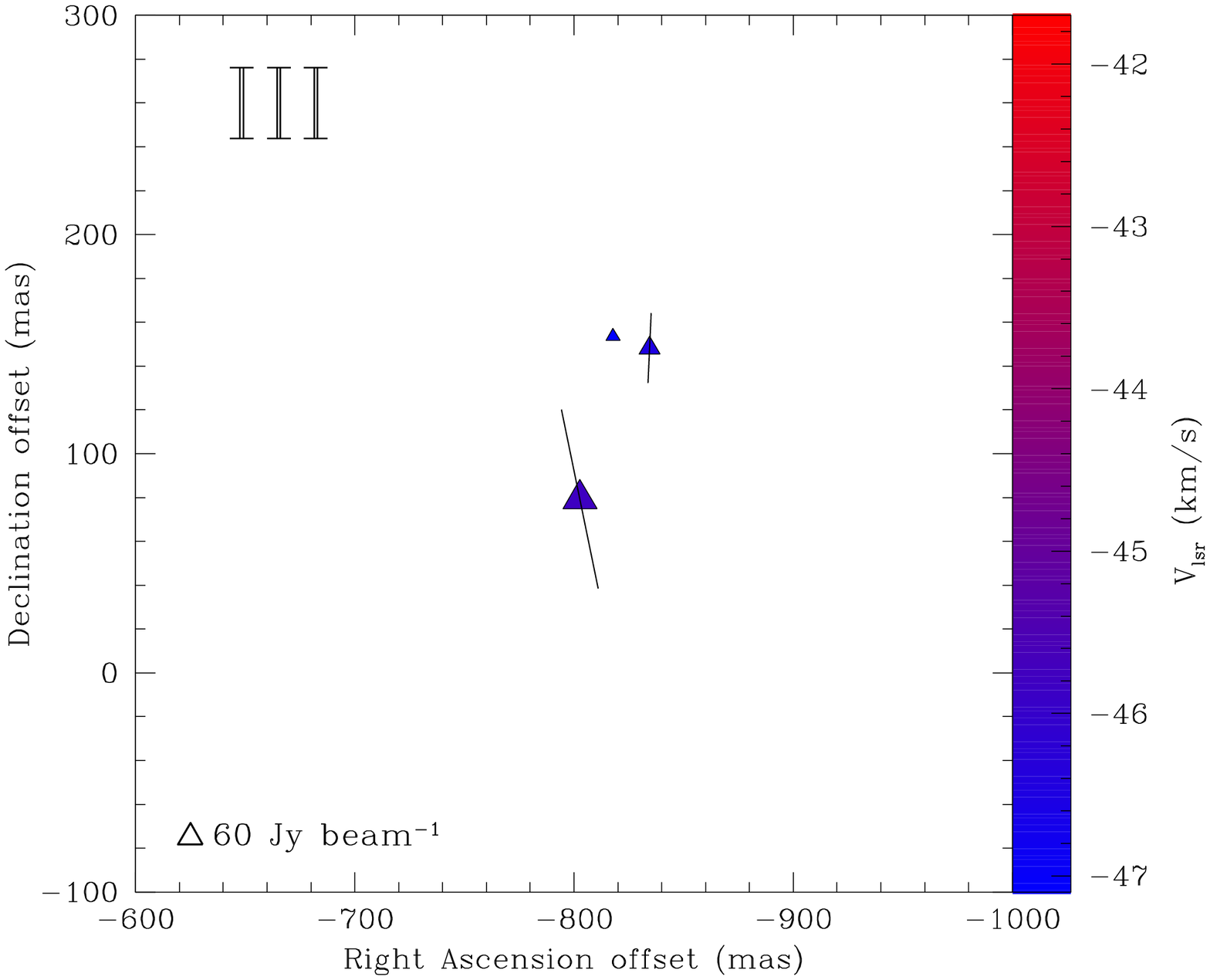}
\includegraphics[width = 7 cm]{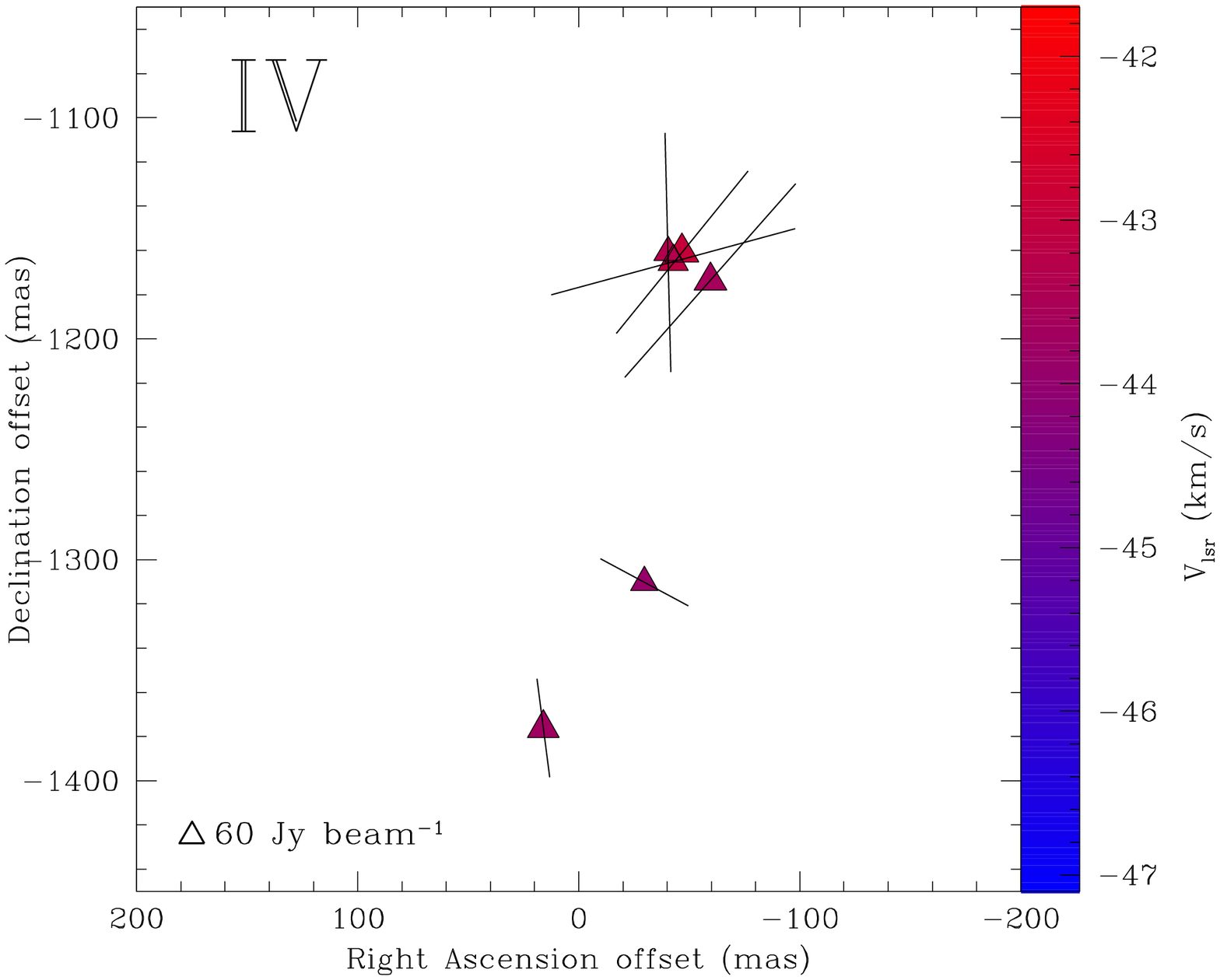}
\includegraphics[width = 7 cm]{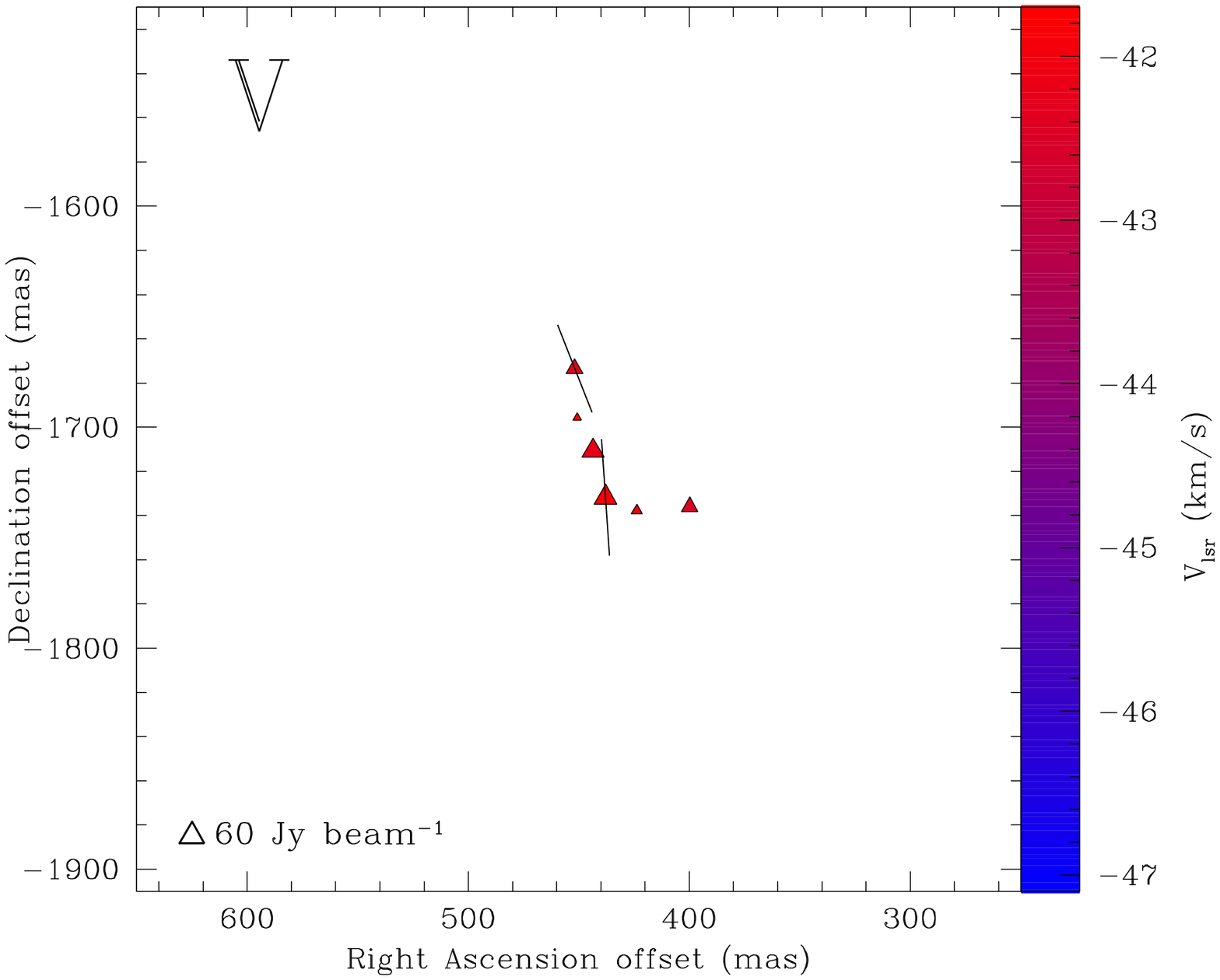}
\includegraphics[width = 7 cm]{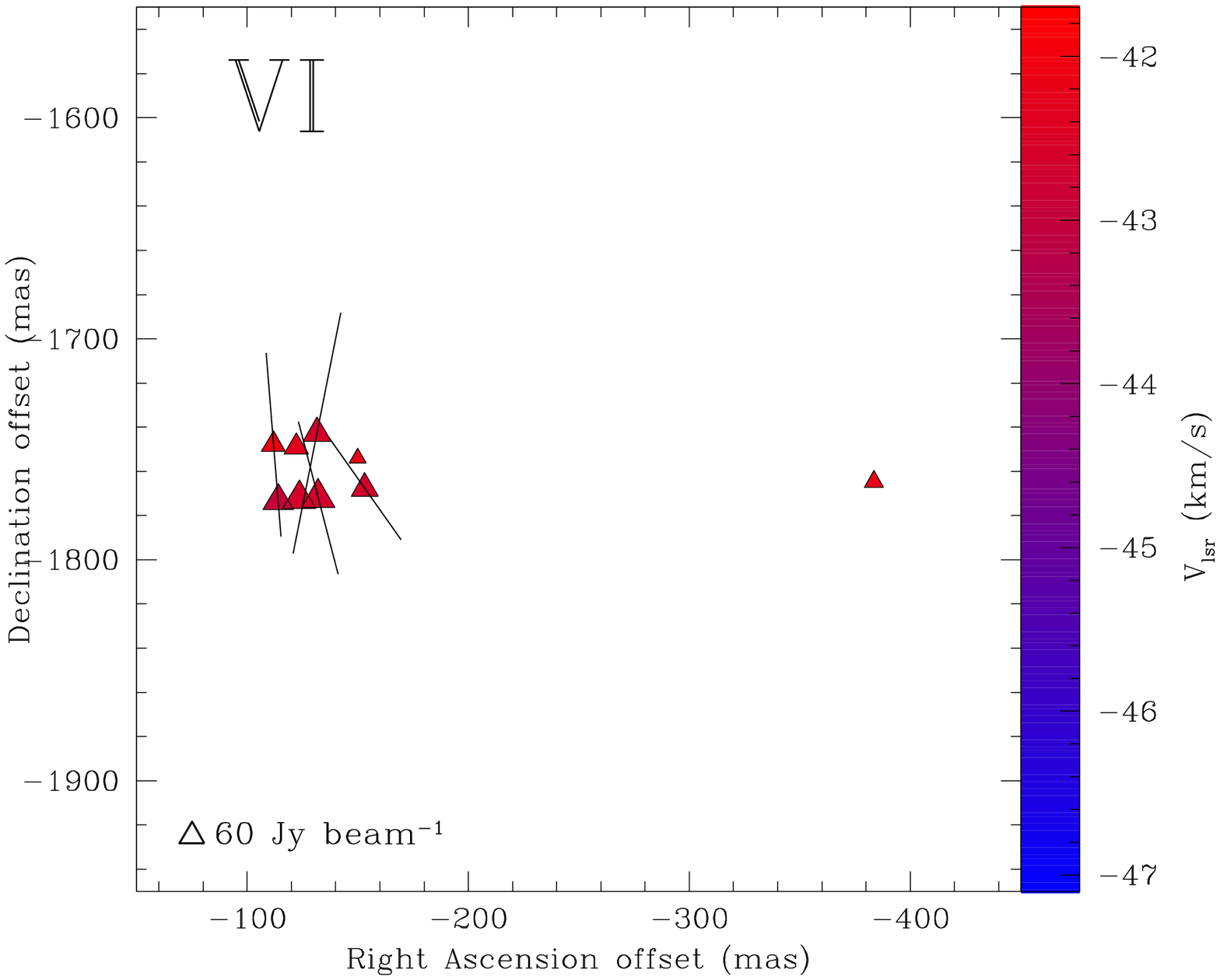}
\caption{Zoom-in view of the six \meth ~maser groups of W3(OH). The dashed line in panel II is the best linear fit of the central \meth ~maser 
features (PA$=121$\d). A, B, and C are used to label three associations of 6.0-GHz OH, 12-GHz \meth, and 6.7-GHz \meth ~masers.}
\label{W3OHreg}
\end{figure*}

W3(OH) is the massive star-forming region towards which we detected one third of all the \meth ~maser features of the First EVN Group. They are listed
in Table~\ref{W3OH_tab}, where they are named as W3OH.01--W3OH.51, and plotted in Fig.~\ref{W3OHcp}. They can be divided in six groups (I--VI) and
their zoom-in views are shown in Fig.~\ref{W3OHreg}. Morphologically the 6.7-GHz \meth ~maser features detected with the EVN show the same filamentary
 structure observed with MERLIN by Harvey-Smith \& Cohen (\cite{har06}) and no new group of features was observed. 
 We do not detected the \meth ~maser features located West of the isolated feature W3OH.05 (features 22 and 23 in Harvey-Smith \& Cohen \cite{har06}).
The linear distribution of the central 6.7-GHz \meth ~maser features of group II (PA$=121$\d) is similar to those
of the 12-GHz \meth ~masers (PA$=141$\d; Moscadelli et al. \cite{mos10}) and of the OH masers (PA$=130$\d$-140$\d; Moscadelli et al. \cite{mos10}), which indicates 
that all the masers complement each other (Moscadelli et al. \cite{mos10}). In the top-right panel of Fig.~\ref{W3OHreg} the three 
associations A, B, and C of 6.0-GHz OH, 12-GHz \meth, and 6.7-GHz \meth ~masers are also reported (Moscadelli et al. \cite{mos10}). Group IV seems
to lie in an arch-like structure. The northern clump (groups I, II, and III) shows velocities in absolute value greater of $\sim2-4$~\kms ~than the 
southern clump (groups IV, V, and VI), in particular group V has in absolute value the lowest velocity ($V_{\rm{lsr}}\approx-42$~\kms).
Along the linear structure of group II the distribution of the maser velocities is well ordered, the most blue-shifted velocities are northwestern
and the most red-shifted are southeastern. This velocity distribution coincides perfectly with that of the 12-GHz \meth ~masers (Moscadelli et al. \cite{mos10}).\\
\indent We have detected linear polarization in more than 50\% of the \meth ~maser features (27/51) with linear polarization fraction ranging from
1.2\% to 8.1\%. The highest $P_{\rm{l}}$ corresponds to the brightest maser feature (W3OH.22) that also shows the highest emerging brightness 
temperature. This value indicates that the feature is partially saturated like 5 other 
features (W3OH.05, W3OH.10, W3OH.14, W3OH.17, and W3OH.18) that have $T_{\rm{b}}\Delta\Omega>2.6\times 10^9$~K~sr. The error weighted emerging brightness 
temperature and the weighted intrinsic thermal linewidth of the unsaturated features are 
$\langle T_{\rm{b}}\Delta\Omega\rangle_{\rm{W3(OH)}}\approx2\times10^{9}$~K~sr and 
$\langle\Delta V_{\rm{i}}\rangle_{\rm{W3(OH)}}=0.9^{+0.4}_{-0.3}$~\kms, respectively. The error weighted linear 
polarization angle is $\langle\chi\rangle_{\rm{W3(OH)}}=3$\d$\pm39$\d, while considering only the unsaturated features 
$\langle\chi\rangle_{\rm{W3(OH)}}=15$\d$\pm31$\d. The $\theta$ values reported in column 13 of 
Table~\ref{W3OH_tab} indicate that for some \meth ~maser features (e.g., W3OH.35) the magnetic field is more likely perpendicular to the linear 
polarization vectors, while for other features it is parallel (e.g., W3OH.09). For 7 \meth ~maser features was possible to 
measure the Zeeman-splitting of which values are in the range $-3.5$~\ms$<\Delta V_{\rm{z}}<3.8$~\ms ~(column 12 of Table~\ref{W3OH_tab}).
\section{Discussion}
\subsection{\meth ~maser properties}
\begin{figure}[h!]
\centering
\includegraphics[width = 8 cm]{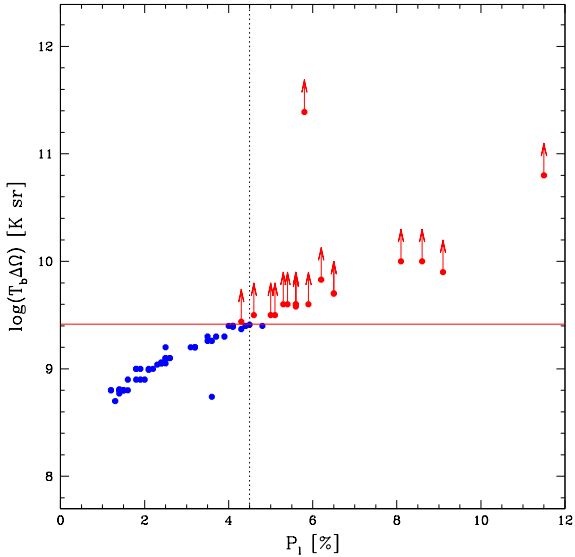}
\caption{The emerging brightness temperatures ($T_{\rm{b}}\Delta\Omega$) as function of the linear polarization fraction ($P_{\rm{l}}$). The blue 
and red circles indicate the unsaturated and saturated masers, respectively, detected in NGC7538 (Surcis et al. \cite{sur11a}), W51, W48, IRAS\,18556+0138, and W3(OH). 
The red arrows indicate that the $T_{\rm{b}}\Delta\Omega$ values obtained from the radiative transfer method code are lower limits. The red full 
line is the limit of emerging brightness temperature above which the \meth ~masers are considered saturated 
($T_{\rm{b}}\Delta\Omega>2.6\times10^9$~K~sr; Surcis et al. \cite{sur11a}), and the dotted line gives the lower limit to the linear polarization fraction for 
saturated masers ($P_{\rm{l}}\approx4.5\%$).}
\label{pltb}
\end{figure}

\begin{figure}[h!]
\centering
\includegraphics[width = 8 cm]{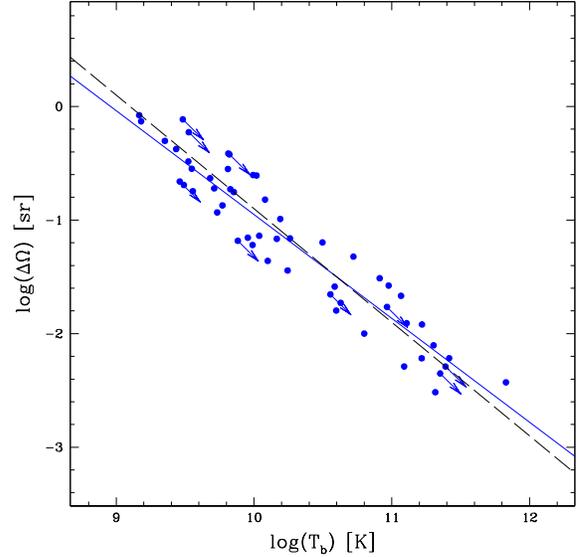}
\caption{The beaming angle ($\Delta\Omega$) as function of the brightness temperatures ($T_{\rm{b}}$) for all the
unsaturated \meth ~masers detected in NGC7538 (Surcis et al. \cite{sur11a}), W51, W48, IRAS\,18556+0138, and W3(OH). The arrows indicate that the
 masers are unresolved, i.e.
the maser angular size $\Sigma$ is less than the beam-size of the observations and consequently $\Delta\Omega$ is smaller and $T_{\rm{b}}$
is larger. The blue line is the best linear fit of the resolved masers and the dashed black line is the theoretical proportionality 
$\Delta\Omega\propto(T_{\rm{b}})^{-1}$ (Elitzur \cite{eli92}).}
\label{omgtb}
\end{figure}

Before discussing the magnetic fields, it is worthwhile examining the properties of the \meth ~maser features (or 
simply \meth~masers). As reported
in Sect.~\ref{res7} we observed a number of saturated \meth ~masers in our EVN group. If we consider all the 72 \meth ~masers detected 
in this group and in NGC7538 for which was possible to determine the emerging brightness temperature (Surcis et al. \cite{sur11a}), we can estimate 
the upper limit of linear polarization fraction for unsaturated 6.7-GHz \meth 
~masers. In Fig~\ref{pltb} we report the emerging brightness temperature as function of the linear polarization fraction. From this plot
we find that the 6.7-GHz \meth ~masers with $P_{\rm{l}}\lesssim4.5\%$ are all unsaturated.
 Furthermore, from the maser theory the beaming angle ($\Delta\Omega$) of unsaturated masers is roughly proportional to 
$|\tau_{\rm{\nu}}|^{-1}$, which is proportional to the brightness temperature $T_{\rm{b}}$ (Elitzur \cite{eli92}). Consequently, if these masers are unsaturated 
$\Delta\Omega\propto(T_{\rm{b}})^{-1}$.  As shown in Surcis et al. (\cite{sur11a}), we can estimate $\Delta\Omega$ by 
comparing the brightness temperature $T_{\rm{b}}$, which is given by %
\begin{equation}
 \frac{T_{\rm{b}}}{[\rm{K}]}=\frac{S(\nu)}{[\rm{Jy}]}\cdot\left(\frac{\Sigma^{2}}{[\rm{mas}^2]}\right)^{-1} \cdot \xi_{\rm{CH_3OH}},
\label{eq71}
\end{equation}
with $T_{\rm{b}}\Delta\Omega$ obtained from the model. In Eq.~\ref{eq71} $S(\nu)$ is the flux density, $\Sigma$ the maser angular diameter, 
 and $\xi_{\rm{CH_3OH}}=13.63 \cdot 10^{9} \rm{~mas^{2} ~Jy^{-1} ~K}$. From the Gaussian fit of the masers $\Sigma$ is between 4 and 8~mas 
and almost all the \meth ~masers appear resolved.
We find that $\Delta\Omega$ ranges from $3\cdot10^{-3}~\rm{sr}$ to $8\cdot10^{-2}$~sr. In Fig.~\ref{omgtb} all the values of
 $\Delta\Omega$ as function of $T_{\rm{b}}$ are reported, both in logarithmic scales. From a linear fit (blue line) we found that
$\Delta\Omega\propto(T_{\rm{b}})^{-0.9}$ implying that the \meth ~masers we have taken into account are indeed unsaturated.\\
\subsection{Magnetic fields in the First EVN Group}
\subsubsection{The importance of magnetic fields in dynamics of massive star formation}
\indent We cannot determine the exact magnetic field strengths from our Zeeman-splitting measurements because of the uncertainty 
of the Land\'{e} g-factor of the 6.7-GHz \meth ~transition from which the Zeeman-splitting coefficient $\alpha_{\rm{Z}}$ is evaluated
 (Vlemmings et al. \cite{vle11}). If we knew the magnetic field strengths we could estimate the importance of the magnetic fields by 
evaluating the ratio between thermal and magnetic energies ($\beta$). If $\beta<1$, the magnetic field dominates the energies in 
high-density protostellar environment.\\
\indent Despite the uncertainty of the magnetic field strengths affects the $\beta$ values we can still
qualitatively determine if the magnetic fields is important in the dynamics of the massive star-forming regions. Substituting 
$\Delta V_{\rm{Z}}=\alpha_{\rm{Z}}\cdot B_{||}$ in Eq.~12 of Surcis et al. (\cite{sur11a}), where $B_{||}$ is the magnetic field 
strength along the line of sight, and
by assuming $n_{\rm{H_{2}}}=10^{9}~\rm{cm^{-3}}$ and $T_{\rm{K}}\sim200$~K, we obtain
\begin{equation}
 \beta=611.6~\alpha_Z^{2} ~cos~\langle\theta\rangle ~\left(\frac{|\Delta V_{\rm{Z}}|}{[\rm{ms^{-1}}]}\right)^{-2}.
\label{eq73}
\end{equation}
Surcis et al. (\cite{sur11a}) argued that a reasonable value for $\alpha_{\rm{Z}}$ should be in the range 
$0.005$~\kmsg~$<\alpha_{\rm{Z}}<0.05$~\kmsg, consequently from Eq.~\ref{eq73} we can estimate a range of $\beta$.\\
\indent The three massive star-forming regions that host \meth ~masers that show circular polarization (W51, W48, and W3OH) have $\beta$
values between $10^{-4}$ and $10^{-1}$. This indicates that the magnetic fields play an important role in all the three regions despite
we cannot exactly determine how much it is because of the uncertainty of Land\'{e} g-factor.
\begin{figure*}[ht!]
\centering
\begin{minipage}{8 cm}
\includegraphics[angle=-90,width = 8 cm]{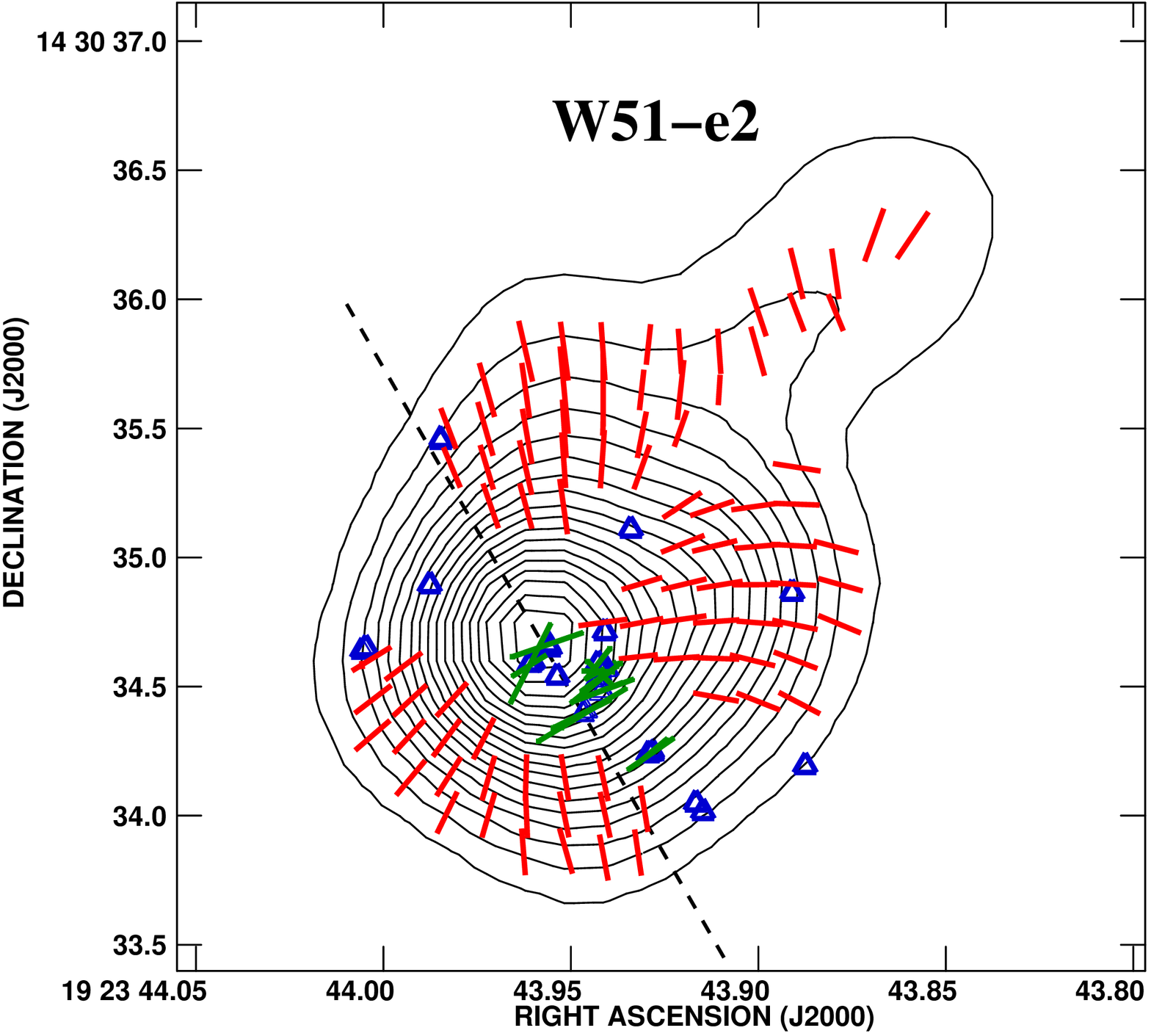}
\end{minipage}
\hfil
\begin{minipage}{10 cm}
\includegraphics[angle=-90,width = 10 cm]{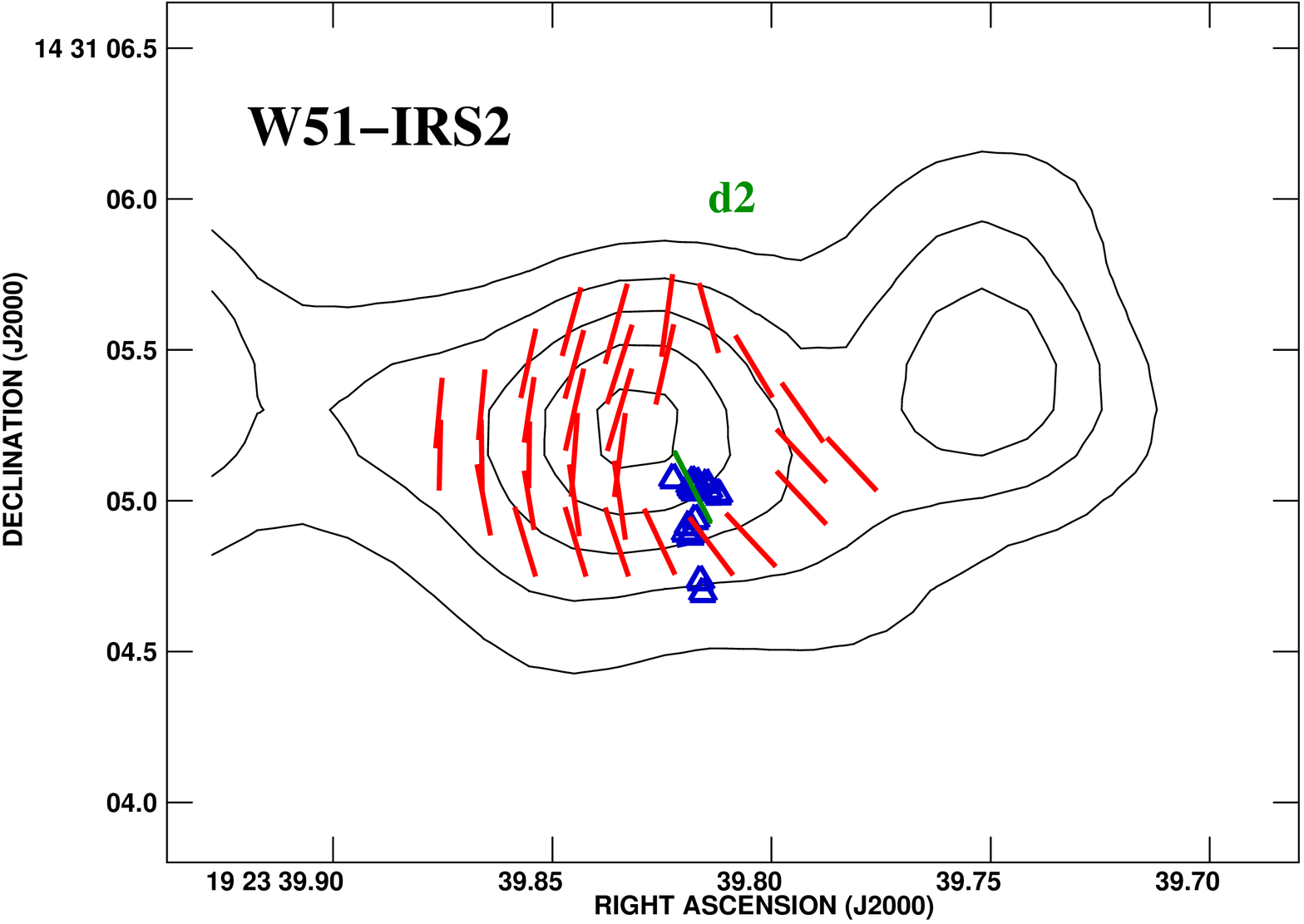}
\end{minipage}
\caption{Left: modified version of Fig.~5a of Tang et al. (\cite{tan09}). The magnetic field (red segments) detected with the SMA (angular resolution $0''\!.7$
that at 5.4~kpc corresponds to $\sim$4000~AU) is 
superimposed on the 870~$\rm{\mu}$m continuum contour map of W51--e2. The dashed black line indicates the direction of the ionized accreting 
flow by Keto \& Klaassen (\cite{ket08}). Right: The magnetic field (red segments) detected with the SMA is 
superimposed on the 870~$\rm{\mu}$m continuum contour map of W51--e2 (Tang et al. private communication). In both images the green segments, which are scaled 
logarithmically according to polarization fraction $P_{\rm{l}}$, mark the direction
 of the magnetic fields as derived from the linearly polarized emission of the \meth ~masers (blue triangles; angular resolution $0''\!.001$ corresponding to
 $\sim5$~AU). }
\label{W51E_MF}
\end{figure*}
\subsubsection{Magnetic field orientations}
The measurements of the linear polarization angles ($\chi$) might be disturbed by the rotation of the linear polarization vectors due
to the medium between the source and the observer. This rotation is the foreground Faraday rotation given by Eq.~6 of Surcis et al. (\cite{sur11a}).
Assuming the interstellar electron density and the magnetic field are $n_{\rm{e}}\approx0.012\,\rm{cm^{-3}}$ and $B_{||}\approx2\,\rm{\mu G}$ 
(Sun et al. \cite{sun08}), $\Phi_{\rm{f}}$ for the 6.7-GHz \meth ~masers can be written as:
\begin{equation}
\Phi_{\rm{f}}[^{\circ}]=2.26~\left(\frac{D}{[\rm{kpc}]}\right),
\label{eq72}
\end{equation}
i.e. the farther the maser source is, the larger $\Phi_{\rm{f}}$ is. This means that the foreground Faraday rotation is estimated to lie between 
$\Phi_{\rm{f}}^{\rm{W51}}\backsimeq12$\d ~and $\Phi_{\rm{f}}^{\rm{W3(OH)}}\backsimeq4$\d.\\
\indent The quite perfect alignment of the linear polarization vectors of the \meth ~masers in W51--e2 suggests that the internal Faraday 
rotation is negligible. Whereas the alignments of the vectors in W48 and IRAS\,18556+0318 additionally implies that the saturation state of the 
masers does not affect the linearly polarized emission. 
Indeed the internal Faraday rotation can destroy the linear polarization (Fish \& Reid \cite{fis06}) but we measured a quite 
high linear polarization fraction ($\sim$5\%) in all the saturated \meth ~masers in W48 and IRAS\,18556+0318. Hence, the contribution of the 
internal Faraday rotation must be negligible both in unsaturated and in saturated \meth ~masers as suggested by the alignments of the linear
polarization vectors.\\

\noindent \textbf{W51}\\
Because $\langle\theta\rangle_{\rm{W51-e2}}=79$\d~$>\theta_{\rm{crit}}$ the 
magnetic field around W51--e2 is on the plane of the sky with a position angle of $\sim120$\d ~that is perfectly perpendicular to the 
ionized accretion flow (PA$\sim$30\d; Keto \& Klaassen \cite{ket08}) rather than to the linear distribution.
The linearly polarized emission has been detected close to the 870~$\rm{\mu}$m continuum peak and so the \meth ~masers are probing 
the magnetic field at the centre of the hourglass morphology mapped by Tang et al. (\cite{tan09}) with the SMA. In Fig.~\ref{W51E_MF} we 
superimposed the results obtained by Tang et al. (\cite{tan09}) with an angular resolution $0''\!.7$ ($\sim$4000~AU) and by us ($\sim5.4$~AU).
 Here we can see that the magnetic field derived from the \meth ~masers 
(green segments) is consistent with the hourglass morphology (red segments). The negative value of the Zeeman-splitting indicates that the magnetic 
field is oriented towards the observer.\\
\indent In the case of W51--IRS2 $\theta_{\rm{W51-e2}}=61^{+11}_{-51}$\d ~indicating that the magnetic field is more likely parallel to the
 linear polarization vectors than perpendicular. Consequently we have also here a perfect agreement with the magnetic field detected
with the SMA at 870~$\rm{\mu}$m by Tang et al. (private communication; Fig.~\ref{W51E_MF}), implying a well-ordered magnetic field around
d2.\\

\noindent \textbf{W48}\\
A comparison between the magnetic fields derived from the 6.7-GHz \meth ~maser observations and from the SCUBA 850~$\rm{\mu}$m observations 
(Curran et al. \cite{cur04}) shows that, like in W51, the magnetic fields are consistent towards the core at different angular resolution.
The positive Zeeman-splitting ($|\Delta V_{\rm{Z}}|_{\rm{W48}}=3.8$~\ms) indicates that the magnetic field is oriented away 
from the observer.\\

\noindent \textbf{IRAS\,18556+0318}\\
Four masers of group A show linearly polarized emission from which the magnetic field is estimated to be oriented East-West (PA~=~90\d).
However, because of the velocity difference among the masers it is worthwhile to determine the orientation of the magnetic field separately by considering
the masers two by two. Considering I18556.04 and I18556.11, which have velocities 7~\kms ~greater than I18556.01 and I18556.08, 
the magnetic field is estimated to have 
PA=112\d, while from the other two masers we estimated PA=83\d. From group B we get PA=80\d. In all cases the magnetic field derived from the 6.7-GHz \meth ~masers 
is aligned neither with the outflow nor with the linear structure, but is almost perpendicular to the radio/infrared jet. \\

\subsubsection{W3(OH)}
Since we observed at higher angular resolution than Vlemmings et al. (\cite{vle06b}) we are able to better determine the magnetic field
 morphology in group II. Comparing Fig.2 of Vlemmings et al. (\cite{vle06b}) and our Fig.\ref{W3OHcp} we see that the linear polarization vectors
 of the 6.7-GHz \meth
 ~masers are not as similar through the whole filamentary structure as observed with the MERLIN, probably because of the 90\d-flip phenomenon. 
However, they appear well ordered within each group (Fig.~\ref{W3OHreg}) and this again indicate that the saturation state
of the masers does not influence the linearly polarized emission.\\
\indent Groups III, V, and VI show error weighted linear polarization angles 
$\langle\chi\rangle$ close to 0\d. A comparison with the results obtained by Vlemmings et al. (\cite{vle06b}), who did not detect any
 linearly polarized emission
 from group V, indicates that the orientation of the linear polarization vectors in these groups are likely influenced by the 90\d-flip 
phenomenon, in particular for group III. Althought the 90\d-flip seems to be confirmed from the $\theta$ angles measured by us for all the three groups
(column 14 of Table~\ref{W3OH_tab}), in groups V and VI the magnetic field is more likely perpendicular to the vectors, i.e. with orientation West-East. Whereas
in group III the magnetic field has an orientation North-South. In group IV we see that all the features but 
\begin{figure}[h!]
\centering
\includegraphics[width = 7 cm]{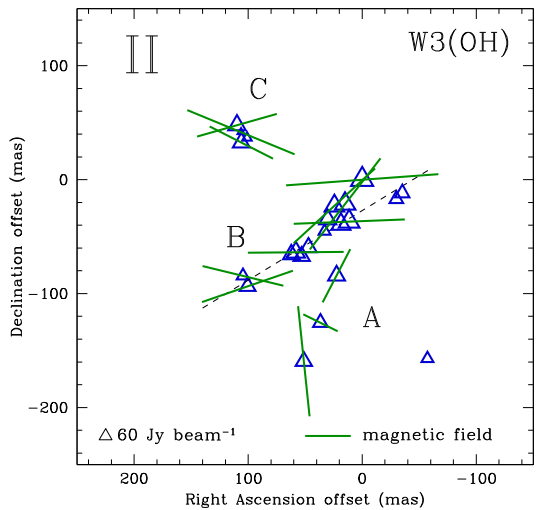}
\caption{Modified version of panel II of Fig.~\ref{W3OHreg}. The green segments, which are scaled 
logarithmically according to polarization fraction $P_{\rm{l}}$, mark the direction of the magnetic field as derived from the 
linearly polarized emission of the \meth ~masers in the northern clump of W3(OH). }
\label{magn_II}
\end{figure}
W3OH.18 and W3OH.25 show a linear polarization angle similar to that measured by Vlemmings et al. (\cite{vle06b}). The different angles of 
W3OH.18 and W3OH.25 could also be due to the 90\d-flip and consequently we find that here, like in group III, the magnetic field is oriented almost North-South. 
This orientation of the magnetic field is also determined from the isolated feature W3OH.05 of which linear polarization angle agrees quite 
perfectly with that measured with MERLIN. While the magnetic field of groups III and VI has an orientation along the filamentary
structure, the magnetic field morphology in group II is completely different.\\
\indent For all the \meth ~masers of the group II-B that have $\chi$ 
close to 0\d ~(W3OH.22, W3OH.23, W3OH.35, W3OH.37, and W3OH.39) we find, considering the errors, that their $\theta$ angles are likely
greater than 55\d, indicating that the magnetic field is perpendicular to the large-scale filamentary structure and almost aligned with the linear distribution
 of the B association. This orientation of the magnetic field is also suggested by the masers that have linear polarization vectors parallel to the linear
 structure (W3OH.24 and W3OH.28) that show $\theta_{\rm{min}}< 55$\d, i.e. magnetic field is likely parallel to them. As show in Fig.~\ref{magn_II}, the \meth ~
masers of A and C associations
 (W3OH.27, W3OH.31, W3OH.41, W3OH.42, and W3OH.43) seems to trace a magnetic field that is changing its morphology from a North-South orientation
to a West-East close to the linear structure (B association).\\
\indent From the Zeeman-splitting measurements we can determine the pointing direction of the magnetic fields. Let consider first group II.
Since we measured both positive (W3OH.22) and negative (W3OH.35 and W3OH.37) Zeeman-splittings at the North-West end and at the South-East end of
the linear distribution respectively, the magnetic field must have a counterclockwise direction. This direction is opposite to
the rotation of the linear distribution, which is clockwise as derived from the OH and \meth ~masers velocities. If the masers along the
linear distribution trace a disc that is rotating clockwise ($M\approx1.5$~\solmass ~and $R\approx270$~AU; Moscadelli et al. \cite{mos10}),
 we have in W3(OH) a scenario 
similar to that of NGC7538 (Surcis et al. \cite{sur11a}). The magnetic field is on the disc (in NGC7538 on the torus) rotating in opposite
 direction to the disc. The 
Zeeman-splitting measurements of the other groups suggest that the magnetic field is twisted along the filamentary structure. 
\section{Conclusions}
We observed a group of 4 massive star-forming regions at 6.7-GHz in full polarization spectral mode with the EVN to detect linear and
circular polarization emission from the \meth ~masers. We detected a total of 154 6.7-GHz \meth ~masers one third of which only towards W3(OH).
We detected linearly polarized emission in all the regions of the EVN group, and in 3 of them we also measured Zeeman-splitting. From our 
observations we have found that the magnetic field is well ordered in all the massive star-forming regions. In W51-e2 the magnetic field derived from the
\meth ~masers is at the centre and along the axis of symmetry of the hourglass morphology observed with the SMA at 870~$\rm{\mu}$m. In W3(OH)
we determined two different magnetic fields, one twisted along the filamentary structure and the other one in the northern clump
on the likely low-mass rotating disc probed by the linear distribution of the \meth ~and OH masers with a counterclockwise orientation. 
In the case of IRAS18556+0318 
(G35.20--0.74N) the magnetic field seems to be perpendicular to the radio/infrared jet rather than along the outflow or along  
a possible torus structure. The magnetic fields are in perfect agreement with their morphology at lower
angular resolution determined with the SMA and SCUBA (W51--e2, W51--d2, and W48) and  they play an important role in all the massive star-forming regions 
of the First EVN Group.\\
\indent In addition, we have also estimated a limit of linear polarization fraction of 4.5\% below which the 6.7-GHz \meth ~masers are 
unsaturated.\\

\noindent \small{\textit{Acknowledgments.} We wish to thank an anonymous referee for making useful suggestions that have improved the paper.
GS and WHTV acknowledge support by the Deutsche Forschungsgemeinschaft (DFG) through the Emmy 
Noether Reseach grant VL 61/3-1. GS thanks Y.-W. Tang for providing the SMA continuum and polarization maps of W51-e1/e2 and W51-North.}

\onltab{1}{
\begin {table*}[t!]
\caption []{All 6.7-GHz methanol maser features detected in W51--e2.} 
\begin{center}
\scriptsize
\begin{tabular}{ l c c c c c c c c c c c }
\hline
\hline
\,\,\,\,\,(1)&(2)& (3)      & (4)          & (5)            & (6)                 & (7)         & (8)      & (9)                     & (10)                        & (11)                   & (12)        \\
Maser & RA\tablefootmark{a} & Dec\tablefootmark{a}& Peak flux    & $V_{\rm{lsr}}$ & $\Delta v\rm{_{L}}$ &$P_{\rm{l}}$ &  $\chi$  & $\Delta V_{\rm{i}}$\tablefootmark{b} & $T_{\rm{b}}\Delta\Omega$\tablefootmark{b} & $\Delta V_{\rm{Z}}$\tablefootmark{c} & $\theta
$\tablefootmark{d}\\
      & offset   & offset   & Density(I)   &                &                     &             &	   &                         &                             &                        &           \\ 
      &  (mas)   & (mas)    & (Jy/beam)       &  (km/s)        &      (km/s)         & (\%)        &   (\d)   & (km/s)                  & (log K sr)                  & (m/s)          & (\d)       \\ 
\hline
W51E.01  & -791.826 & -357.414 &$0.78\pm0.02$ &  51.54        &     0.24             &$-$          &  $-$     &  $-$                    & $-$                         & $-$	      &  $-$ \\ 
W51E.02  & -741.361 & 313.950  &$0.24\pm0.01$ &  53.03        &     0.22             &$-$          &  $-$     &  $-$                    & $-$                         & $-$	      &  $-$ \\ 
W51E.03  & -396.633 & -535.208 &$2.31\pm0.02$ &  51.45        &     0.42             &$-$          &  $-$     &  $-$                    & $-$                         & $-$	      &  $-$ \\ 
W51E.04  & -368.770 & -501.720 &$0.47\pm0.01$ &  52.42        &     0.22             &$-$          &  $-$     &  $-$                    & $-$                         & $-$	      &  $-$ \\ 
W51E.05  & -202.860 & -304.199 &$1.11\pm0.04$ &  51.10        &     0.33             &$-$          &  $-$     &  $-$                    & $-$                         & $-$	      &  $-$ \\ 
W51E.06  & -198.317 & -300.964 &$9.73\pm0.04$ &  51.80        &     0.27             &$1.6\pm0.2$  & $40\pm12$&  $1.3^{+0.1}_{-0.2}$    & $8.9^{+0.6}_{-0.2}$         & $-$	      &  $79^{+10}_{-40}$ \\ 
W51E.07  & -193.719 & -310.570 &$5.45\pm0.03$ &  52.15        &     0.27             &$2.2\pm0.3$  & $35\pm49$&  $1.2^{+0.1}_{-0.2}$    & $9.0^{+0.6}_{-0.2}$         & $-$	      &  $80^{+10}_{-39}$ \\ 
W51E.08  & -185.520 & -311.840 &$0.13\pm0.01$ &  52.94        &     0.33             &$-$          &  $-$     &  $-$                    & $-$                         & $-$	      &  $-$ \\ 
W51E.09  & -116.608 & 558.552  &$3.84\pm0.16$ &  59.35        &     0.21             &$-$          & $-$      &  $-$                    & $-$                         & $-$	      &  $-$ \\ 
W51E.10  & -20.773  & 16.247   &$2.05\pm0.06$ &  60.14        &     0.17             &$-$          &  $-$     &  $-$                    & $-$                         & $-$	      &  $-$ \\ 
W51E.11  & -16.176  & 160.938  &$2.02\pm0.05$ &  59.88        &     0.25             &$-$          & $-$      &  $-$                    & $-$                         & $-$	      &  $-$ \\ 
W51E.12  & -7.368   & -65.472  &$22.72\pm0.03$& 58.30         &     0.27             &$2.6\pm0.6$  & $20\pm6$ &  $0.8^{+0.1}_{-0.1}$    & $9.1^{+0.5}_{-0.6}$         & $-$	      &  $79^{+11}_{-17}$ \\ 
W51E.13  & -2.936   & -65.970  &$65.78\pm0.20$& 57.86         &     0.36             &$1.8\pm0.4$  & $21\pm5$ &  $1.7^{+0.1}_{-0.2}$    & $8.9^{+0.7}_{-0.5}$         & $-0.4\pm0.1$     &  $77^{+13}_{-37}$ \\ 
W51E.14  & 0        & 0        &$217.32\pm0.23$& 59.26        &     0.33             &$1.4\pm0.8$  & $46\pm9$ &  $1.7^{+0.1}_{-0.2}$    & $8.8^{+0.4}_{-1.3}$         & $-0.7\pm0.1$     &  $67^{+11}_{-47}$ \\ 
W51E.15  & 0.443    & 12.135   &$26.63\pm0.07$&  59.00        &     0.25             &$1.4\pm0.4$  & $29\pm18$&  $1.2^{+0.1}_{-0.1}$    & $8.8^{+0.6}_{-0.7}$         & $-$	      &  $74^{+12}_{-41}$ \\ 
W51E.16  & 0.831    & -76.960  &$2.28\pm0.03$ &  58.39        &     0.27             &$-$          & $-$      &  $-$                    & $-$                         & $-$	      &  $-$ \\ 
W51E.17  & 1.219    & -14.587  &$50.54\pm0.05$&  59.97        &     0.38             &$1.9\pm0.4$  & $35\pm15$&  $1.7^{+0.6}_{-0.6}$    & $8.9^{+0.1}_{-0.5}$         & $-$	      &  $90^{+11}_{-11}$ \\ 
W51E.18  & 4.819    & 9.281    &$27.56\pm0.16$&  59.35        &     0.28             &$1.2\pm0.2$  & $-2\pm32$&  $-$                    & $-$                         & $-2.4\pm0.1$     &  $-$ \\ 
W51E.19  & 10.581   & 38.508   &$4.09\pm0.05$ &  59.97        &     0.27             &$1.5\pm1.2$  & $52\pm17$&  $0.7^{+0.1}_{-0.1}$    & $8.8^{+0.4}_{-1.0}$         & $-$              &  $68^{+12}_{-46}$ \\ 
W51E.20  & 11.412   & -24.912  &$2.94\pm0.05$ &  60.05        &     0.25             &$-$          &  $-$     &  $-$                    & $-$                         & $-$	      &  $-$ \\ 
W51E.21  & 27.975   & -43.551  &$3.15\pm0.01$ &  60.32        &     0.23             &$2.3\pm1.6$  & $39\pm46$&  $-$                    & $-$                         & $-$	      &  $-$ \\ 
W51E.22  & 56.393   & -143.511 &$2.09\pm0.04$ &  58.91        &     0.19             &$-$          &  $-$     &  $-$                    & $-$                         & $-$	      &  $-$ \\ 
W51E.23  & 67.250   & -145.042 &$15.37\pm0.05$& 58.74         &     0.28             &$2.0\pm0.5$  & $29\pm11$&  $1.1^{+0.1}_{-0.1}$    & $8.9^{+0.5}_{-0.7}$         & $-$	      &  $78^{+12}_{-37}$ \\ 
W51E.24  & 69.300   & -158.505 &$9.48\pm0.07$ &  59.00        &     0.30             &$4.0\pm1.6$  & $31\pm6$ &  $0.6^{+0.1}_{-0.1}$    & $9.4^{+0.2}_{-1.3}$         & $-$	      &  $84^{+6}_{-42}$ \\ 
W51E.25  & 70.851   & -154.001 &$8.19\pm0.03$ &  58.82        &     0.26             &$2.5\pm1.1$  & $26\pm8$ &  $0.7^{+0.1}_{-0.1}$    & $9.1^{+0.3}_{-1.1}$         & $-$	      &  $76^{+14}_{-40}$ \\ 
W51E.26  & 168.514  & -12.121  &$4.27\pm0.03$ &  58.56        &     0.51             &$-$          &  $-$     &  $-$                    & $-$                         & $-$	      &  $-$ \\ 
W51E.27  & 196.655  & 99.159   &$1.82\pm0.05$ &  59.80        &     0.23             &$-$          &  $-$     &  $-$                    & $-$                         & $-$	      &  $-$ \\ 
W51E.28  & 204.632  & 113.516  &$5.67\pm0.02$ &  60.58        &     0.34             &$2.9\pm0.6$  &  $19\pm8$&  $-$                    & $-$                         & $-$	      &  $-$ \\ 
W51E.29  & 262.520  & 43.556   &$2.26\pm0.07$ &  58.91        &     0.29             &$-$          &  $-$     &  $-$                    & $-$                         & $-$	      &  $-$ \\ 
W51E.30  & 268.004  & 38.809   &$1.99\pm0.03$ &  58.82        &     0.27             &$-$          &  $-$     &  $-$                    & $-$                         & $-$	      &  $-$ \\ 
W51E.31  & 271.716  & 40.205   &$3.30\pm0.03$ &  58.47        &     0.40             &$3.5\pm1.1$  &  $63\pm5$&  $1.4^{+0.1}_{-0.2}$    & $9.3^{+0.2}_{-1.5}$         & $-$	      &  $84^{+6}_{-40}$ \\ 
W51E.32  & 276.535  & 31.513   &$8.11\pm0.06$ &  57.86        &     0.48             &$1.6\pm0.5$  & $32\pm23$&  $0.6^{+0.1}_{-0.1}$    & $8.8^{+0.5}_{-0.6}$         & $-2.7\pm0.1$     &  $75^{+14}_{-35}$ \\ 
W51E.33  & 616.831  & 909.365  &$0.57\pm0.07$ &  51.01        &     0.25             &$-$          &  $-$     &  $-$                    & $-$                         & $-$	      &  $-$ \\ 
W51E.34  & 620.709  & 902.697  &$0.66\pm0.02$ &  50.04        &     0.40             &$-$          &  $-$     &  $-$                    & $-$                         & $-$	      &  $-$ \\ 
W51E.35  & 661.979  & 343.477  &$0.62\pm0.03$ &  58.47        &     0.22             &$-$          &  $-$     &  $-$                    & $-$                         & $-$	      &  $-$ \\ 
W51E.36  & 909.542  & 98.515   &$0.69\pm0.03$ &  58.82        &     0.28             &$-$          &  $-$     &  $-$                    & $-$                         & $-$	      &  $-$ \\ 
W51E.37  & 923.834  & 88.762   &$1.29\pm0.04$ &  58.47        &     0.30             &$-$          &  $-$     &  $-$                    & $-$                         & $-$	      &  $-$ \\ 
\hline
\end{tabular}
\end{center}
\tablefoot{
\tablefoottext{a}{The reference position is $\alpha_{2000}=19^{\rm{h}}23^{\rm{m}}43^{\rm{s}}\!.942$ and $\delta_{2000}=+14^{\circ}30'34''\!\!.550$.}
\tablefoottext{b}{The best-fitting results obtained by using a model based on the radiative transfer theory of methanol masers 
for $\Gamma+\Gamma_{\nu}=1$ (Vlemmings et al. \cite{vle10}, Surcis et al. \cite{sur11b}). The errors were determined by analyzing the full probability 
distribution function.}
\tablefoottext{c}{The Zeeman-splittings are determined from the cross-correlation between the RR and LL spectra.}
\tablefoottext{d}{ The angle between the magnetic field and the maser propagation direction is determined by using the observed $P_{\rm{l}}$ 
and the fitted emerging brightness temperature. The errors were determined by analyzing the full probability distribution function.}
}
\label{WE_tab}
\end{table*}
}
\onltab{2}{
\begin {table*}[t!]
\caption []{All 6.7-GHz methanol maser features detected in W51--IRS2.} 
\begin{center}
\scriptsize
\begin{tabular}{ l c c c c c c c c c c c c }
\hline
\hline
\,\,\,\,\,(1)&(2)& (3)      & (4)          & (5)            & (6)                 & (7)         & (8)      & (9)                     & (10)                        & (11)                   & (12)            \\
Maser & RA\tablefootmark{a} & Dec\tablefootmark{a}& Peak flux    & $V_{\rm{lsr}}$ & $\Delta v\rm{_{L}}$ &$P_{\rm{l}}$ &  $\chi$  & $\Delta V_{\rm{i}}
$\tablefootmark{b} & $T_{\rm{b}}\Delta\Omega$\tablefootmark{b}& $\Delta V_{\rm{Z}}$\tablefootmark{c}&  $\theta$\tablefootmark{d}\\
      & offset   & offset   & Density(I)   &                &                     &             &	   &                         &                             &                        &            \\ 
      &  (mas)   & (mas)    & (Jy/beam)       &  (km/s)        &      (km/s)         & (\%)        &   (\d)   & (km/s)                  & (log K sr)                  & (m/s)          &  (\d)       \\ 
\hline
W51N.01  & -85.860  & -28.289  &$0.27\pm0.01$ &  54.52         &     0.31            &$-$          &  $-$     &  $-$                    & $-$                         & $-$	            & $-$ \\ 
W51N.02  & -83.367  & -22.108  &$0.20\pm0.01$ &  54.96         &     0.28            &$-$          &  $-$     &  $-$                    & $-$                         & $-$	     	    & $-$ \\ 
W51N.03  & -53.067  & -11.479  &$0.33\pm0.04$ &  57.94         &     0.28            &$-$          &  $-$     &  $-$                    & $-$                         & $-$	      	    & $-$ \\ 
W51N.04  & -45.312  &  8.416   &$0.90\pm0.02$ &  57.51         &     0.25            &$-$          &  $-$     &  $-$                    & $-$                         & $-$	            & $-$ \\ 
W51N.05  & -33.569  & -351.122 &$0.61\pm0.02$ &  59.70         &     0.18            &$-$          &  $-$     &  $-$                    & $-$                         & $-$	            & $-$ \\ 
W51N.06  & -25.536  & -312.023 &$0.47\pm0.03$ &  59.88         &     0.28            &$-$          &  $-$     &  $-$                    & $-$                         & $-$	            & $-$ \\ 
W51N.07  & -12.962  &  13.928  &$0.74\pm0.02$ &  55.84         &     0.25            &$-$          &  $-$     &  $-$                    & $-$                         & $-$	            & $-$ \\ 
W51N.08  & -9.140   & -108.629 &$1.02\pm0.03$ &  56.28         &     0.29            &$-$          &  $-$     &  $-$                    & $-$                         & $-$	            & $-$ \\ 
W51N.09  & -6.813   & -2.965   &$1.93\pm0.01$ &  56.45         &     0.21            &$-$          &  $-$     &  $-$                    & $-$                         & $-$	            & $-$ \\ 
W51N.10  & -5.152   & -16.856  &$1.12\pm0.03$ &  56.28         &     0.35            &$-$          &  $-$     &  $-$                    & $-$                         & $-$	            & $-$ \\ 
W51N.11  & 0        & 0        &$12.03\pm0.07$&  56.19         &     0.35            &$1.3\pm1.1$  & $27\pm13$&  $1.8^{+0.1}_{-0.1}$    & $8.7^{+0.2}_{-1.4}$         & $-$	            & $61^{+11}_{-51}$ \\ 
W51N.12  & 1.496    & 18.406   &$0.95\pm0.01$ &  56.63         &     0.36            &$-$          &  $-$     &  $-$                    & $-$                         & $-$	            & $-$ \\ 
W51N.13  & 5.096    & -6.713   &$1.25\pm0.01$ &  55.75         &     0.30            &$-$          &  $-$     &  $-$                    & $-$                         & $-$	            & $-$ \\ 
W51N.14  & 5.373    & -161.138 &$2.42\pm0.05$ &  59.79         &     0.24            &$-$          &  $-$     &  $-$                    & $-$                         & $-$	            & $-$ \\ 
W51N.15  & 10.192   & 6.065    &$0.57\pm0.01$ &  56.98         &     0.31            &$-$          &  $-$     &  $-$                    & $-$                         & $-$	            & $-$ \\ 
W51N.16  & 12.630   & -129.178 &$0.48\pm0.02$ &  60.84         &     0.26            &$-$          &  $-$     &  $-$                    & $-$                         & $-$	            & $-$ \\ 
W51N.17  & 24.761   & -149.938 &$0.16\pm0.01$ &  60.23         &     0.26            &$-$          &  $-$     &  $-$                    & $-$                         & $-$	            & $-$ \\ 
W51N.18  & 64.644   & 26.281   &$0.20\pm0.02$ &  55.75         &     0.28            &$-$          &  $-$     &  $-$                    & $-$                         & $-$	            & $-$ \\ 
\hline
\end{tabular}
\end{center}
\tablefoot{
\tablefoottext{a}{The reference position is $\alpha_{2000}=19^{\rm{h}}23^{\rm{m}}39^{\rm{s}}\!.818$ and
 $\delta_{2000}=+14^{\circ}31'5''\!\!.045$.}
\tablefoottext{b}{The best-fitting results obtained by using a model based on the radiative transfer theory of methanol masers 
 for $\Gamma+\Gamma_{\nu}=1$ (Vlemmings et al. \cite{vle10}, Surcis et al. \cite{sur11b}). The errors were determined by 
analyzing the full probability distribution function.}
\tablefoottext{c}{The Zeeman-splittings are determined from the cross-correlation between the RR and LL spectra.}
\tablefoottext{d}{The angle between the magnetic field and the maser propagation direction is determined by using the observed $P_{\rm{l}}$ 
and the fitted emerging brightness temperature. The errors were determined by analyzing the full probability distribution function.}
}
\label{WN_tab}
\end{table*}
}
\onltab{3}{
\begin{table*}[t]
\caption []{All 6.7 GHz methanol maser features detected in W48.} 
\begin{center}
\scriptsize
\begin{tabular}{ l c c c c c c c c c c c c }
\hline
\hline
\,\,\,\,\,(1)&(2)& (3)      & (4)          & (5)            & (6)                 & (7)         & (8)      & (9)                     & (10)                        & (11)                   & (12)          \\
Maser & RA       & Dec      & Peak flux    & $V_{\rm{lsr}}$ & $\Delta v\rm{_{L}}$ &$P_{\rm{l}}$ &  $\chi$  & $\Delta V_{\rm{i}}$\tablefootmark{a} & $T_{\rm{b}}\Delta\Omega$\tablefootmark{a}& $\Delta V_{\rm{Z}}$\tablefootmark{b}&  $\theta$\tablefootmark{c}\\
      & offset   & offset   & Density(I)   &                &                     &             &	   &                         &                             &                        &            \\ 
      &  (mas)   & (mas)    & (Jy/beam)       &  (km/s)        &      (km/s)         & (\%)        &   (\d)   & (km/s)                  & (log K sr)                  & (m/s)          &  (\d)       \\ 
\hline
W48.01& -138.785 & -87.692  &$19.00\pm0.40$&  43.78         &   0.22              &$-$          &  $-$     &  $-$                    & $-$                         & $-$	            & $-$ \\ 
W48.02& -110.296 & -57.672  &$6.84\pm0.06$ &  44.13         &   0.19              &$-$          &  $-$     &  $-$                    & $-$                         & $-$	            & $-$ \\ 
W48.03& -109.953 & -128.784 &$14.07\pm0.26$&  41.41         &   0.23              &$-$          &  $-$     &  $-$                    & $-$                         & $-$	            & $-$ \\ 
W48.04\tablefootmark{d}& -101.143 & -70.511  &$60.05\pm1.46$&  42.91         &   0.29              &$5.1\pm0.2$  &  $23\pm8$&  $0.6^{+0.2}_{-0.1}$    & $9.5^{+0.3}_{-0.1}$         & $-$	            & $87^{+3}_{-13}$ \\ 
W48.05& -99.026  & -134.666 &$23.28\pm0.95$&  40.89         &   0.29              &$-$          &  $-$     &  $-$                    & $-$                         & $-$	            & $-$ \\ 
W48.06& -94.621  & -117.296 &$8.38\pm0.14$ &  42.03         &   0.24              &$-$          &  $-$     &  $-$                    & $-$                         & $-$	            & $-$ \\ 
W48.07& -90.502  & -158.745 &$78.93\pm1.40$&  41.50         &   0.22              &$-$          &  $-$     &  $-$                    & $-$                         & $-$	            & $-$ \\ 
W48.08& -89.072  & -140.818 &$1.58\pm0.02$ &  40.18         &   0.35              &$-$          &  $-$     &  $-$                    & $-$                         & $-$	            & $-$ \\ 
W48.09& -82.093  & -128.763 &$3.82\pm0.09$ &  40.18         &   0.35              &$-$          &  $-$     &  $-$                    & $-$                         & $-$	            & $-$ \\ 
W48.10& -81.635  & -122.286 &$202.87\pm6.56$& 42.47         &   0.26              &$4.4\pm1.3$  & $22\pm6$ &  $0.7^{+0.1}_{-0.1}$    & $9.4^{+0.2}_{-1.4}$         & $-$	            & $84^{+6}_{-20}$\\ 
W48.11& -81.235  & -157.660 &$47.04\pm0.08$&  40.97         &   0.31              &$-$          &  $-$     &  $-$                    & $-$                         & $-$	            & $-$ \\ 
W48.12& -77.974  & -200.914 &$22.79\pm0.18$&  41.76         &   0.37              &$-$          &  $-$     &  $-$                    & $-$                         & $-$	            & $-$ \\ 
W48.13& -75.171  & -192.154 &$11.58\pm0.26$&  41.41         &   0.17              &$-$          &  $-$     &  $-$                    & $-$                         & $-$	            & $-$ \\ 
W48.14\tablefootmark{d}& 0        &  0       &$294.68\pm1.11$& 44.49         &   0.24              &$4.6\pm1.5$  & $12\pm6$ &  $1.0^{+0.1}_{-0.2}$    & $9.5^{+0.6}_{-1.1}$         & $3.8\pm0.4$            & $73^{+16}_{-32}$ \\ 
W48.15& 13.158   & -65.594  &$2.48\pm0.12$ &  43.34         &   0.15              &$-$          &  $-$     &  $-$                    & $-$                         & $-$	            & $-$ \\ 
W48.16& 27.631   & -0.439   &$26.15\pm0.41$&  44.57         &   0.26              &$-$          &  $-$     &  $-$                    & $-$                         & $-$	            & $-$ \\ 
W48.17& 27.631   & -29.171  &$22.64\pm1.11$&  44.49         &   0.12              &$-$          &  $-$     &  $-$                    & $-$                         & $-$	            & $-$ \\ 
W48.18& 34.267   & -20.596  &$3.05\pm0.17$ &  45.36         &   0.24              &$-$          &  $-$     &  $-$                    & $-$                         & $-$	            & $-$ \\ 
W48.19\tablefootmark{d}& 47.082   & -62.460  &$101.52\pm1.30$& 43.70         &   0.25              &$5.4\pm0.6$  & $16\pm4$ &  $0.7^{+0.2}_{-0.1}$    & $9.6^{+0.4}_{-0.2}$         & $-$	            & $84^{+4}_{-12}$ \\ 
W48.20\tablefootmark{d}& 58.352   & 6.796    &$32.74\pm0.23$&  44.84         &   0.25              &$6.5\pm0.6$  & $27\pm1$ &  $0.6^{+0.2}_{-0.1}$    & $9.7^{+0.3}_{-0.2}$         & $-$	            & $86^{+3}_{-10}$ \\ 
\hline
\end{tabular}
\end{center}
\tablefoot{
\tablefoottext{a}{The best-fitting results obtained by using a model based on the radiative transfer theory of methanol masers 
for $\Gamma+\Gamma_{\nu}=1$ (Vlemmings et al. \cite{vle10}, Surcis et al. \cite{sur11b}). The errors were determined by analyzing the full probability 
distribution function.}
\tablefoottext{b}{The Zeeman-splittings are determined from the cross-correlation between the RR and LL spectra.}
\tablefoottext{c}{The angle between the magnetic field and the maser propagation direction is determined by using the observed $P_{\rm{l}}$ 
and the fitted emerging brightness temperature. The errors were determined by analyzing the full probability distribution function.}
\tablefoottext{d}{Because of the degree of the saturation of these \meth ~masers $T_{\rm{b}}\Delta\Omega$ is underestimated, $\Delta V_{\rm{i}}$ 
and $\theta$ are overestimated.}
}
\label{W48_tab}
\end{table*}
}
\onltab{4}{
\begin {table*}[]
\caption []{All 6.7 GHz methanol maser features detected in IRAS18556+01 (G35.20--0.74N).} 
\begin{center}
\scriptsize
\begin{tabular}{ l c c c c c c c c c c c c c }
\hline
\hline
\,\,\,\,\,(1)&(2)& (3)      & (4)          & (5)            & (6)                 & (7)         & (8)      & (9)                     & (10)                        & (11)                   & (12)     &(13)     \\
Maser & Group & RA      & Dec      & Peak flux   & $V_{\rm{lsr}}$ & $\Delta v\rm{_{L}}$ &$P_{\rm{l}}$ &  $\chi$  & $\Delta V_{\rm{i}}$\tablefootmark{a} & $T_{\rm{b}}\Delta\Omega$\tablefootmark{a}& $\Delta V_{\rm{Z}}$\tablefootmark{b}&  $\theta$\tablefootmark{c}\\
      &       & offset  & offset  & Density(I)   &                &                     &             &	   &                         &                             &                        &            \\ 
      &       & (mas)   & (mas)    & (Jy/beam)   &  (km/s)        &      (km/s)         & (\%)        &   (\d)   & (km/s)                  & (log K sr)                  & (m/s)          &  (\d)       \\ 
\hline
I18556.01\tablefootmark{d}  & A     & -1587.159& 1817.554 &$14.71\pm0.34$&    29.89       &    0.23             &$8.6\pm1.3$  & $-9\pm3$ &  $<0.5$                 & $10.0^{+0.3}_{-0.9} $       & $-$	            & $90^{+14}_{-14} $ \\ 
I18556.02  & A     & -1563.866& 1764.340 &$1.05\pm0.10$ &    28.57       &    0.42             &$-$          &  $-$     &  $-$                    & $-$                         & $-$	            & $-$ \\ 
I18556.03  & A     & -1561.521& 1881.611 &$0.85\pm0.03$ &    33.22       &    0.30             &$-$          &  $-$     &  $-$                    & $-$                         & $-$	            & $-$ \\ 
I18556.04\tablefootmark{d}  & A     & -1558.818& 1630.295 &$4.66\pm0.03$ &    36.56       &    0.26             &$9.1\pm1.4$  &$22\pm28$ &  $<0.5$                 & $9.9^{+0.2}_{-1.2} $        & $-$	            & $90^{+11}_{-11}$ \\ 
I18556.05  & A     & -1559.276& 1826.489 &$4.47\pm0.03$ &    27.95       &    0.24             &$-$          &  $-$     &  $-$                    & $-$                         & $-$	            & $-$ \\ 
I18556.06  & A     & -1553.628& 1790.443 &$0.32\pm0.03$ &    27.95       &    0.26             &$-$          &  $-$     &  $-$                    & $-$                         & $-$	            & $-$ \\ 
I18556.07  & A     & -1551.154& 1729.794 &$0.69\pm0.02$ &    27.60       &    0.28             &$-$          &  $-$     &  $-$                    & $-$                         & $-$	            & $-$ \\ 
I18556.08\tablefootmark{d}  & A     & -1550.882& 1862.656 &$12.13\pm0.08$&    29.36       &    0.29             &$6.5\pm0.7$  &$2\pm3$   &  $1.0^{+0.2}_{-0.1} $   & $9.7^{+0.4}_{-0.2} $        & $-$	            & $84^{+4}_{-11} $ \\ 
I18556.09  & A     & -1549.567& 1674.839 &$1.31\pm0.02$ &    30.41       &    0.30             &$-$          &  $-$     &  $-$                    & $-$                         & $-$	            & $-$ \\ 
I18556.10  & A     & -1548.166& 1827.927 &$0.52\pm0.02$ &    27.25       &    0.22             &$-$          &  $-$     &  $-$                    & $-$                         & $-$	            & $-$ \\ 
I18556.11  & A     & -1548.080& 1604.614 &$12.81\pm0.11$&    35.42       &    0.30             &$4.8\pm1.1$  &$22\pm18$ &  $0.7^{+0.2}_{-0.1} $   & $9.4^{+0.4}_{-1.2} $        & $-$	            & $84^{+6}_{-15}$ \\ 
I18556.12  & A     & -1538.571& 1715.225 &$0.14\pm0.01$ &    33.49       &    0.30             &$-$          &  $-$     &  $-$                    & $-$                         & $-$	            & $-$ \\ 
I18556.13  & A     & -1535.311& 1903.782 &$0.97\pm0.03$ &    33.22       &    0.20             &$-$          &  $-$     &  $-$                    & $-$                         & $-$	            & $-$ \\ 
I18556.14  & A     & -1527.890& 1870.255 &$0.56\pm0.02$ &    33.40       &    0.14             &$-$          &  $-$     &  $-$                    & $-$                         & $-$	            & $-$ \\ 
I18556.15  & A     & -1519.568& 1904.572 &$1.74\pm0.01$ &    26.90       &    0.43             &$-$          &  $-$     &  $-$                    & $-$                         & $-$	            & $-$ \\ 
I18556.16  & A     & -1513.805& 1912.239 &$1.36\pm0.01$ &    26.20       &    0.31             &$-$          &  $-$     &  $-$                    & $-$                         & $-$	            & $-$ \\ 
I18556.17  & A     & -1501.437& 1774.460 &$0.22\pm0.01$ &    30.94       &    0.21             &$-$          &  $-$     &  $-$                    & $-$                         & $-$	            & $-$ \\ 
I18556.18  & A     & -1466.290& 1754.551 &$0.20\pm0.01$ &    32.34       &    0.37             &$-$          &  $-$     &  $-$                    & $-$                         & $-$	            & $-$ \\ 
I18556.19  & A     & -1454.021& 1765.602 &$0.14\pm0.01$ &    31.73       &    0.27             &$-$          &  $-$     &  $-$                    & $-$                         & $-$	            & $-$ \\ 
I18556.20  & A     & -1293.200& 1994.118 &$4.37\pm0.19$ &    29.10       &    0.21             &$-$          &  $-$     &  $-$                    & $-$                         & $-$	            & $-$ \\ 
I18556.21  & B     & -14.056  & 20.645   &$1.41\pm0.02$ &    27.95       &    0.40             &$-$          &  $-$     &  $-$                    & $-$                         & $-$	            & $-$ \\ 
I18556.22  & B     & -14.013  & 28.847   &$3.86\pm0.05$ &    28.31       &    0.23             &$-$          &  $-$     &  $-$                    & $-$                         & $-$	            & $-$ \\ 
I18556.23  & B     & -12.669  & 72.609   &$1.45\pm0.05$ &    28.31       &    0.65             &$-$          &  $-$     &  $-$                    & $-$                         & $-$	            & $-$ \\ 
I18556.24  & B     & -11.039  & 17.769   &$0.39\pm0.01$ &    27.43       &    0.25             &$-$          &  $-$     &  $-$                    & $-$                         & $-$	            & $-$ \\ 
I18556.25  & B     & -5.863   & -4.211   &$0.31\pm0.01$ &    27.34       &    0.19             &$-$          &  $-$     &  $-$                    & $-$                         & $-$	            & $-$ \\ 
I18556.26  & B     & -3.117   & -277.435 &$1.58\pm0.05$ &    28.39       &    0.21             &$-$          &  $-$     &  $-$                    & $-$                         & $-$	            & $-$ \\ 
I18556.27\tablefootmark{d}& B     & 0        & 0        &$104.81\pm1.05$&   28.57       &    0.26             &$11.5\pm1.9$ & $-10\pm5$&  $<0.5 $                & $10.8^{+0.5}_{-0.7}$        & $-$	            & $85^{+5}_{-7}$ \\ 
I18556.28  & B     & 49.389   & 48.145   &$0.89\pm0.03$ &    27.69       &    0.18             &$-$          &  $-$     &  $-$                    & $-$                         & $-$	            & $-$ \\ 
\hline
\end{tabular}
\end{center}
\tablefoot{
\tablefoottext{a}{The best-fitting results obtained by using a model based on the radiative transfer theory of methanol masers 
for $\Gamma+\Gamma_{\nu}=1$ (Vlemmings et al. \cite{vle10}, Surcis et al. \cite{sur11b}). The errors were determined by analyzing the full probability 
distribution function.}
\tablefoottext{b}{The Zeeman-splittings are determined from the cross-correlation between the RR and LL spectra.}
\tablefoottext{c}{The angle between the magnetic field and the maser propagation direction is determined by using the observed $P_{\rm{l}}$ 
and the fitted emerging brightness temperature. The errors were determined by analyzing the full probability distribution function.}
\tablefoottext{d}{Because of the degree of the saturation of these \meth ~masers $T_{\rm{b}}\Delta\Omega$ is underestimated, $\Delta V_{\rm{i}}$ 
and $\theta$ are overestimated.}
}
\label{IRAS_tab}
\end{table*}
}
\onltab{5}{
\begin {table*}[t!]
\caption []{All 6.7-GHz methanol maser features detected in W3(OH).} 
\begin{center}
\scriptsize
\begin{tabular}{ l c c c c c c c c c c c c c }
\hline
\hline
\,\,\,\,\,(1)&(2)& (3)      & (4)          & (5)            & (6)                 & (7)         & (8)      & (9)                     & (10)                        & (11)                   & (12)     &(13)     \\
Maser & Group & RA\tablefootmark{a} & Dec\tablefootmark{a}& Peak flux    & $V_{\rm{lsr}}$ & $\Delta v\rm{_{L}}$ &$P_{\rm{l}}$ &  $\chi$  & $\Delta V_{\rm{i}}$\tablefootmark{b} & $T_{\rm{b}}\Delta\Omega$\tablefootmark{b} & $\Delta V_{\rm{Z}}$\tablefootmark{c}&  $\theta$\tablefootmark{d}\\
      &       & offset   & offset   & Density(I)   &                &                     &             &	   &                         &                             &                        &            \\ 
      &       & (mas)   & (mas)    & (Jy/beam)       &  (km/s)        &      (km/s)         & (\%)        &   (\d)   & (km/s)                  & (log K sr)                  & (m/s)          &  (\d)       \\ 
\hline
W3OH.01 & III   & -834.479 & 148.296  &$11.58\pm0.02$  &   -46.55       &    0.20             &$1.2\pm0.4$  & $-2\pm10$&  $0.9^{+0.1}_{-0.1}$    & $8.8^{+0.6}_{-0.5}$         & $-$	            & $71^{+10}_{-42}$ \\ 
W3OH.02 & III   & -817.800 & 153.654  &$2.26\pm0.01$   &   -47.07       &    0.18             &$-$          &  $-$     &  $-$                    & $-$                         & $-$	            & $-$ \\ 
W3OH.03 & III   & -802.674 & 79.334   &$27.78\pm0.28$  &   -45.76       &    0.22             &$3.2\pm0.8$  & $12\pm69$&  $1.1^{+0.1}_{-0.2}$    & $9.2^{+0.8}_{-0.4}$         & $-$	            & $69^{+9}_{-42}$ \\ 
W3OH.04 & VI    & -383.492 & -1764.656&$7.78\pm0.04$   &   -42.24       &    0.19             &$-$          &  $-$     &  $-$                    & $-$                         & $-$	            & $-$ \\ 
W3OH.05\tablefootmark{e}& - &-162.497 & -691.895 &$33.61\pm0.55$  &   -44.00       &    0.18             &$5.6\pm1.7$  & $-69\pm5$&  $0.7^{+0.1}_{-0.1}$    & $9.6^{+0.4}_{-1.2}$         & $-$	            & $83^{+7}_{-14}$ \\ 
W3OH.06 & VI    & -122.187 & -1749.006&$27.32\pm0.05$  &   -42.33       &    0.21             &$-$          &  $-$     &  $-$          	       & $-$                         & $-$	            & $-$ \\ 
W3OH.07 & VI    & -153.206 & -1767.649&$54.81\pm0.34$  &   -42.60       &    0.20             &$1.9\pm0.6$  & $35\pm14$&  $0.9^{+0.1}_{-0.1}$    & $9.0^{+0.6}_{-0.7}$         & $-$	            & $72^{+18}_{-34}$ \\ 
W3OH.08 & VI    & -149.964 & -1753.939&$4.82\pm0.04$   &   -42.24       &    0.19             &$-$          &  $-$     &  $-$                    & $-$                         & $-$	            & $-$ \\ 
W3OH.09 & VI    & -132.122 & -1771.915&$235.45\pm0.33$ &   -42.60       &    0.19             &$2.5\pm1.2$  & $15\pm33$&  $0.9^{+0.1}_{-0.1}$    & $9.2^{+0.7}_{-0.9}$         & $-$	            & $67^{+16}_{-38}$ \\ 
W3OH.10\tablefootmark{e} &VI&-131.591 & -1742.706&$60.79\pm0.34$  &   -42.60       &    0.18             &$5.3\pm2.7$  & $-11\pm7$&  $0.8^{+0.1}_{-0.1}$    & $9.6^{+0.4}_{-1.3}$         & $-$	            & $77^{+12}_{-42}$ \\ 
W3OH.11 & VI    & -123.541 & -1772.516&$212.03\pm0.33$ &   -42.60       &    0.22             &$-$          &  $-$     &  $-$                    & $-$                         & $-3.5\pm0.2$         & $-$ \\ 
W3OH.12 & VI    & -114.142 & -1773.373&$111.69\pm0.37$ &   -42.95       &    0.20             &$-$          &  $-$     &  $-$                    & $-$                         & $-$	            & $-$ \\ 
W3OH.13 & VI    & -111.926 & -1747.892&$22.93\pm0.03$  &   -42.07       &    0.20             &$3.1\pm0.3$  &  $5\pm38$&  $0.8^{+0.1}_{-0.1}$    & $9.2^{+0.3}_{-0.1}$         & $-$	            & $81^{+8}_{-14}$ \\ 
W3OH.14\tablefootmark{e} &IV&-59.565  & -1173.670&$182.60\pm0.64$ &   -43.65       &    0.26             &$5.9\pm1.9$  &$-41\pm44$&  $1.1^{+0.1}_{-0.2}$    & $9.6^{+0.5}_{-1.3}$         & $-$	            & $75^{+15}_{-29}$ \\ 
W3OH.15 & II    & -57.240  & -157.034 &$20.812\pm0.59$ &   -44.79       &    0.33             &$-$          &  $-$     &  $-$                    & $-$                         & $-$	            & $-$ \\ 
W3OH.16 & IV    & -46.725  & -1160.805&$233.50\pm0.38$ &   -42.95       &    0.32             &$3.9\pm0.5$  &$-39\pm86$&  $1.5^{+0.1}_{-0.3}$    & $9.3^{+0.6}_{-0.2}$         & $-$	            & $74^{+11}_{-39}$ \\ 
W3OH.17\tablefootmark{e} &IV&-42.741  & -1165.085&$110.50\pm0.29$ &   -43.12       &    0.25             &$5.6\pm3.9$  &$-75\pm41$&  $1.1^{+0.2}_{-0.2}$    & $9.6^{+0.1}_{-1.8}$         & $-$	            & $75^{+14}_{-45}$ \\ 
W3OH.18\tablefootmark{e} &IV&-40.408  & -1160.982&$65.18\pm0.49$  &   -43.12       &    0.29             &$5.0\pm2.6$  & $2\pm85$ &  $1.2^{+0.1}_{-0.2}$    & $9.5^{+0.2}_{-1.7}$         & $-$	            & $74^{+13}_{-43}$ \\ 
W3OH.19 & II    & -35.097  & -11.770  &$69.09\pm0.62$  &   -44.79       &    0.25             &$-$          &  $-$     &  $-$                    & $-$                         & $-$	            & $-$ \\ 
W3OH.20 & II    & -30.011  & -17.119  &$46.52\pm0.72$  &   -44.70       &    0.25             &$-$          &  $-$     &  $-$                    & $-$                         & $-$	            & $-$ \\ 
W3OH.21 & IV    & -29.630  & -1310.143&$57.91\pm0.52$  &   -43.91       &    0.31             &$1.5\pm1.1$  & $62\pm49$&  $1.5^{+0.1}_{-0.2}$    & $8.8^{+0.7}_{-0.4}$         & $-$	            & $77^{+13}_{-38}$ \\ 
W3OH.22\tablefootmark{e} &II& 0  & 0  &$2051.30\pm1.42$&   -45.41       &    0.31             &$8.1\pm0.7$  & $4\pm34$ &  $1.0^{+0.1}_{-0.1}$    & $10.0^{+0.5}_{-0.1}$        & $1.9\pm0.1$          & $73^{+10}_{-5}$ \\ 
W3OH.23 & II    & 11.410   & -36.799  &$1099.80\pm0.81$&   -43.39       &    0.29             &$4.1\pm1.2$  &  $2\pm41$&  $1.2^{+0.1}_{-0.2}$    & $9.4^{+0.6}_{-1.0}$         & $-$	            & $75^{+14}_{-36}$ \\ 
W3OH.24 & II    & 14.956   & -21.265  &$913.15\pm0.92$ &   -44.53       &    0.37             &$4.4\pm2.0$  &$-38\pm41$&  $1.70^{+0.03}_{-0.25}$ & $9.4^{+0.9}_{-1.0}$         & $-$	            & $68^{+9}_{-45}$ \\ 
W3OH.25 & IV    & 15.969   & -1376.160&$156.18\pm0.65$ &   -43.74       &    0.21             &$1.5\pm0.4$  &  $7\pm23$&  $1.1^{+0.1}_{-0.1}$    & $8.8^{+0.7}_{-0.4}$         & $3.8\pm0.5$          & $68^{+7}_{-45}$ \\ 
W3OH.26 & II    & 18.653   & -39.053  &$546.64\pm0.63$ &   -43.65       &    0.21             &$-$          &  $-$     &  $-$                    & $-$                         & $-$	            & $-$ \\ 
W3OH.27 & II    & 22.492   & -84.309  &$206.85\pm0.36$ &   -43.03       &    0.18             &$1.8\pm0.2$  & $63\pm35$&  $0.9^{+0.1}_{-0.1}$    & $9.0^{+0.5}_{-0.1}$         & $-$	            & $76^{+5}_{-46}$ \\ 
W3OH.28 & II    & 24.383   & -22.999  &$640.49\pm0.70$ &   -44.70       &    0.29             &$4.1\pm1.6$  &$-47\pm19$&  $1.4^{+0.1}_{-0.2}$    & $9.4^{+0.8}_{-1.0}$         & $-$	            & $68^{+12}_{-40}$ \\ 
W3OH.29 & II    & 30.256   & -34.817  &$239.84\pm0.56$ &   -44.09       &    0.27             &$-$          &  $-$     &  $-$                    & $-$                         & $-$	            & $-$ \\ 
W3OH.30 & II    & 33.002   & -45.343  &$20.09\pm0.63$  &   -43.56       &    0.26             &$-$          &  $-$     &  $-$                    & $-$                         & $-$	            & $-$ \\ 
W3OH.31 & II    & 39.591   & -125.509 &$74.16\pm0.50$  &   -43.30       &    0.19             &$1.2\pm0.9$  & $64\pm29$&  $0.9^{+0.1}_{-0.1}$    & $8.8^{+0.3}_{-1.2}$         & $-$	            & $65^{+7}_{-52}$ \\ 
W3OH.32 & II    & 47.238   & -59.534  &$155.18\pm0.36$ &   -43.03       &    0.18             &$-$          &  $-$     &  $-$                    & $-$                         & $-$	            & $-$ \\ 
W3OH.33 & II    & 51.173   & -159.398 &$146.66\pm0.36$ &   -43.03       &    0.26             &$4.1\pm2.7$  &  $6\pm42$&  $1.2^{+0.1}_{-0.2}$    & $9.4^{+0.2}_{-1.6}$         & $-$	            & $70^{+13}_{-45}$ \\ 
W3OH.34 & II    & 53.063   & -67.207  &$140.91\pm0.98$ &   -44.53       &    0.40             &$-$          &  $-$     &  $-$                    & $-$                         & $-$	            & $-$ \\ 
W3OH.35 & II    & 58.220   & -63.723  &$347.43\pm0.37$ &   -42.86       &    0.19             &$3.2\pm1.1$  & $1\pm37$ &  $0.9^{+0.1}_{-0.1}$    & $9.2^{+0.6}_{-0.9}$         & $-1.5\pm0.1$         & $71^{+18}_{-33}$ \\ 
W3OH.36 & II    & 62.103   & -65.691  &$180.85\pm0.70$ &   -44.35       &    0.37             &$-$          &  $-$     &  $-$                    & $-$                         & $-$	            & $-$ \\ 
W3OH.37 & II    & 100.540  & -93.575  &$110.11\pm0.28$ &   -42.51       &    0.31             &$3.2\pm0.8$  & $19\pm53$&  $1.4^{+0.1}_{-0.2}$    & $9.2^{+0.7}_{-0.8}$         & $-1.7\pm0.4$         & $76^{+13}_{-37}$ \\ 
W3OH.38 & II    & 101.041  & 36.774   &$7.17\pm0.02$   &   -46.37       &    0.25             &$-$          &  $-$     &  $-$                    & $-$                         & $-$	            & $-$ \\ 
W3OH.39 & II    & 104.515  & -84.400  &$28.12\pm0.05$  &   -42.33       &    0.27             &$2.7\pm1.0$  &  $-14\pm6$& $0.8^{+0.2}_{-0.1}$    & $9.3^{+0.4}_{-1.1}$         & $-$	            & $81^{+9}_{-17}$ \\ 
W3OH.40 & I     & 105.043  & 540.987  &$178.62\pm0.45$ &   -45.14       &    0.22             &$-$          &  $-$     &  $-$                    & $-$                         & $-3.4\pm0.4$         & $-$ \\ 
W3OH.41 & II    & 106.053  & 32.572   &$19.21\pm0.28$  &   -45.76       &    0.31             &$2.1\pm0.3$  & $63\pm80$&  $1.7^{+0.1}_{-0.2}$    & $9.0^{+0.7}_{-0.2}$         & $-$   	            & $69^{+6}_{-46}$ \\ 
W3OH.42 & II    & 106.329  & 41.679   &$2.08\pm0.01$   &   -46.99       &    0.44             &$4.4\pm0.5$  & $68\pm5$ &  $-$                    & $-$                         & $-$	            & $-$ \\ 
W3OH.43 & II    & 109.883  & 47.815   &$254.77\pm0.69$ &   -44.35       &    0.24             &$2.6\pm0.7$  & $16\pm27$&  $1.2^{+0.1}_{-0.1}$    & $9.1^{+0.7}_{-0.7}$         & $-$	            & $73^{+11}_{-41}$ \\ 
W3OH.44 & I     & 117.485  & 535.473  &$73.74\pm0.44$  &   -45.05       &    0.17             &$-$          &  $-$     &  $-$                    & $-$                         & $-$	            & $-$ \\ 
W3OH.45 & I     & 170.642  & 437.187  &$0.72\pm0.01$   &   -47.07       &    0.18             &$-$          &  $-$     &  $-$                    & $-$                         & $-$	            & $-$ \\ 
W3OH.46 & V     & 399.814  & -1735.970&$4.32\pm0.11$   &   -42.42       &    0.43             &$-$          &  $-$     &  $-$                    & $-$                         & $-$	            & $-$ \\ 
W3OH.47 & V     & 423.769  & -1737.766&$1.25\pm0.02$   &   -41.89       &    0.28             &$-$          &  $-$     &  $-$                    & $-$                         & $-$	            & $-$ \\ 
W3OH.48 & V     & 437.887  & -1731.754&$19.97\pm0.02$  &   -41.81       &    0.78             &$1.8\pm0.3$  &  $4\pm2$ &  $1.0^{+0.1}_{-0.1}$    & $9.0^{+0.5}_{-0.3}$         & $1.4\pm0.2$          & $76^{+11}_{-40}$ \\ 
W3OH.49 & V     & 443.434  & -1710.760&$15.71\pm0.04$  &   -42.24       &    0.26             &$-$          &  $-$     &  $-$                    & $-$                         & $-$	            & $-$ \\ 
W3OH.50 & V     & 450.607  & -1695.734&$0.62\pm0.02$   &   -41.72       &    0.27             &$-$          &  $-$     &  $-$                    & $-$                         & $-$	            & $-$ \\ 
W3OH.51 & V     & 451.835  & -1673.563&$4.46\pm0.04$   &   -42.24       &    0.22             &$1.5\pm0.2$  &  $22\pm5$&  $0.55^{+0.09}_{-0.01}$ & $8.8^{+0.4}_{-0.1}$         & $-$	            & $79^{+11}_{-35}$ \\ 
\hline
\end{tabular}
\end{center}
\tablefoot{
\tablefoottext{a}{The reference position is $\alpha_{2000}=02^{\rm{h}}27^{\rm{m}}03^{\rm{s}}\!.833$ and 
$\delta_{2000}=+61^{\circ}52'25''\!\!.288$.}
\tablefoottext{b}{The best-fitting results obtained by using a model based on the radiative transfer theory of methanol masers 
for $\Gamma+\Gamma_{\nu}=1$ (Vlemmings et al. \cite{vle10}, Surcis et al. \cite{sur11b}). The errors were determined 
by analyzing the full probability distribution function.}
\tablefoottext{c}{The Zeeman-splittings are determined from the cross-correlation between the RR and LL spectra.}
\tablefoottext{d}{The angle between the magnetic field and the maser propagation direction is determined by using the observed $P_{\rm{l}}$ 
and the fitted emerging brightness temperature. The errors were determined by analyzing the full probability distribution function.}
\tablefoottext{e}{Because of the degree of the saturation of these \meth ~masers $T_{\rm{b}}\Delta\Omega$ is underestimated, $\Delta V_{\rm{i}}$ 
and $\theta$ are overestimated.}
}
\label{W3OH_tab}
\end{table*}
}

\listofobjects
\end{document}